\documentclass[cup6a]{cupbook}
\usepackage{latexsym,amssymb,amsmath,amsfonts,graphicx,url,times}
\usepackage[PetersLenny]{fncychap}

\newcommand{\HRule}{\hskip -.6cm \rule{\linewidth}{0.2mm}}

\begin{document}




\pagenumbering{arabic}
\setcounter{chapter}{0}


\author[Panos Aliferis]{Panos Aliferis\\Capital Fund Management, 6 blvd Haussmann, 75009 Paris, France \\ {aliferis.panos@gmail.com}}

\chapter{} 

The theory of quantum error correction provides a general methodology for protecting quantum information from noise. It is therefore expected that quantum error correction will be essential in operating future quantum computers, machines storing and manipulating very large amounts of quantum information in the course of long quantum computations.

Certainly, there is nothing particularly quantum mechanical in the idea of encoding the information stored and processed inside computers. Ordinary digital computers already use various {\em fault tolerance} methods at the software level to correct errors during the storage or the transmission of information---e.g., the integrity of the bits stored in hard disks is verified by using parity checks (checksums). In addition, for critical computing systems such as these inside airplanes or nuclear reactors, software fault tolerance methods are also applied during the {\em processing} of information---e.g., airplane control computers compare the results from multiple parallel processors to detect faults. In general however, the hardware of modern digital computers is remarkably robust to noise so that, for most applications, the use of additional software error correction is rather limited. 

In contrast to the easiness and robustness with which classical information can be processed\footnote[1]{Of course, this was not always the case; photographs of ENIAC, the first universal electrical computer, speak volumes about how difficult the first steps of classical computing were.}, the processing of quantum information appears at present to be much more challenging. Although constructing reliable quantum computing hardware is certainly a daunting task, we have nevertheless strong hopes that large-scale quantum computers, able to implement useful long computations, can in fact be realized. This optimism is founded on methods of {\em quantum fault tolerance} which show that scalable quantum computation is, in principle, possible against a variety of noise processes. Demonstrating that these methods work effectively in practice is a major challenge for contemporary science, a challenge whose outcome will depend on our progress in understanding the physical noise processes in experiments, and on our ability to design and optimize fault tolerance methods according to the limitations and the noise characteristics of experimental devices. 

This chapter is an introduction to software methods of quantum fault tolerance. Broadly speaking, these methods describe strategies for using the noisy hardware components of a quantum computer to perform computations while continually monitoring and actively correcting the hardware faults. The methods we will discuss are general and apply independently of how the hardware components are physically realized in the laboratory. Nevertheless, one should not lose sight of the fact that what we describe in this chapter as elementary hardware components are {\em not} elementary from an experimental point of view. Already at the level of the realization of qubits in the laboratory, the experimenter strives to choose implementations with high inherent robustness to noise such as qubits encoded in decoherence-free subspaces or noiseless subsystems, or qubits which are topologically protected. In addition, 
noise in the elementary hardware operations can be suppressed by using various open-loop techniques such as refocusing or dynamical decoupling. Even though these various qubit encodings and noise-suppression techniques can be highly effective, some residual noise will always remain; it is this residual effective noise that needs to be treated by the error correction and fault tolerance methods we will discuss in this chapter. 

The basic conceptual ideas of fault tolerance for quantum computation are very similar as in the case of classical computation: First, a code is chosen and each logical step of the computation is implemented by a fault-tolerant {\em gadget} which acts on the encoded information; these gadgets comprise many elementary hardware operations, and they are designed to implement the desired logical transformation on the encoded information while at the same time detecting and correcting errors. And secondly, the protection from noise is increased by designing a hierarchy of encoding layers such that errors become progressively weaker as we pass from one layer to the next. 


Despite these similarities, there are two major differences between quantum and classical fault tolerance, which are related to the differences between classical and quantum error correction. 
The first difference is that in the quantum case error correction needs to be implemented {\em coherently}, i.e., in a way that preserves the quantum superpositions in the encoded information that is processed by the quantum computer---this requirement has no analogue in classical fault tolerance since quantum superpositions and quantum interference play no role in classical computation. The second difference relates to the types of noise that are of concern in the two cases. For ordinary computers that manipulate classical information digitized in bits, noise can simply be viewed as causing abrupt changes in the value of each bit (bit flips). For quantum computers on the other hand, information is stored in quantum states which (if pure) are in general superpositions $a_1e^{i\phi_1}|\psi_1\rangle + a_2e^{i\phi_2}|\psi_2\rangle + \cdots$ of various physically relevant basis states $|\psi_i\rangle$ with real coefficients $a_i$ and $\phi_i$; then not only can noise cause changes in the amplitudes $a_i$ (which is analogous to the bit flip errors for classical information as $a_i^2$ is the probability of occupation of the state $|\psi_i\rangle$) but noise can also cause changes to the phases $\phi_i$ (phases are irrelevant when storing classical information but they are important quantum mechanically as they determine the ability of the superposed basis states to interfere). 

\section{Quantum circuits and error discretization} \vspace{.3cm}
\label{sec:qc-discretization}


The precise character of noise in future quantum computers will depend on the particular hardware implementation. There is a wide variety of prospective implementation schemes that are being experimentally investigated at present, but in this chapter we restrict the discussion to those schemes which fall under the quantum circuit model for which methods of fault tolerance are better understood\footnote[1]{In particular, we will not discuss quantum computations realized purely by {\em adiabatic} evolution for which a general theory of fault tolerance is lacking. 
}. 

\begin{figure}[t]
\begin{tabular}{c}
\put(-5.8,0){\includegraphics[width=14cm,keepaspectratio]{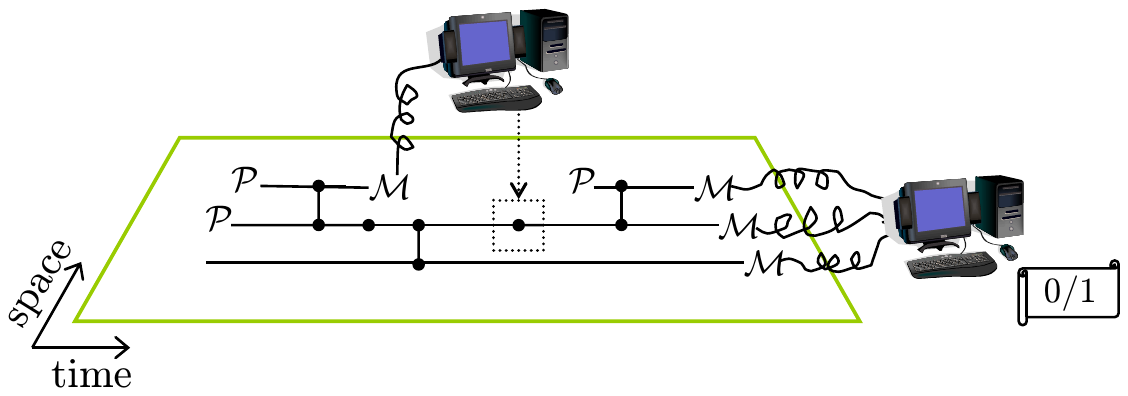}}
\vspace{-6.8cm}
\end{tabular}
\caption{\label{fig:circuits} A quantum circuit is a sequence of qubit preparations ($\mathcal{P}$), qubit measurements ($\mathcal{M}$), and quantum gates (a $\bullet $  denotes a gate applied on a single qubit, and two $\bullet $'s connected vertically denote a gate applied on a pair of qubits). The measurement outcomes are processed by classical computers alongside the quantum computer; intermediate measurement outcomes condition the application of future quantum gates, while the final measurement outcomes encode the answer of the quantum computation (0 or 1).}
\end{figure}

Quantum circuits are a generalization of classical circuits: A classical circuit computing a boolean function $f$ on $n$ bits is a prescription for expressing $f$ as a composition of functions or {\em gates} that act on a fixed, independent of $n$, number of bits at a time; gates are chosen from a finite set which is {\em universal} allowing any function to be computed---e.g., the {\sc not} gate which flips the value of a bit together with the {\sc and} gate which computes the conjunction of the value of two bits form a universal gate set. Similarly, a quantum circuit is a prescription for implementing a physical operation on the Hilbert space $\mathcal{H}_n = (\mathbb{C}^2)^{\otimes n}$ of $n$ qubits as a composition of elementary physical operations which are applied on a fixed number of qubits at a time. Although $\mathcal{H}_n$ is continuous, there exist {\em finite} sets of physical operations acting on at most two qubits which are {\em quantum universal} allowing the approximation of any physical operation in $\mathcal{H}_n$ to any desired accuracy; these universal sets comprise preparations of single qubits in certain pure states, certain unitary transformations or {\em quantum gates} on single qubits or between pairs of qubits, and measurements of single-qubit observables. Our diagrammatic representation of a quantum circuit is shown in fig.~\ref{fig:circuits}.

We will not discuss examples of quantum universal sets here. Conceptually, what is important is that a finite number of elementary operations suffices for implementing any quantum computation. Therefore, a quantum computer is a discrete machine just like a classical digital computer; in the classical case, the elementary hardware components are gates on one or a few bits, while in the quantum case they are physical operations on one or a few qubits. Discreteness is essential both classically and quantumly because it implies that fault tolerance can be achieved by constructing fault-tolerant gadgets for each one of the operations in a universal set and composing these gadgets together. 

The second essential ingredient for fault tolerance is the ability to discretize errors so that error correction becomes possible. In classical digital computation, the basis of error discretization is the digital encoding of information. Once a bit of information is represented in a physical quantity taking the value $v_0$ to encode 0 and $v_1$ to encode 1, noise in gates can be described in terms of discrete errors taking $v_0$ to $v_1$ or vice versa, and small fluctuations around the values $v_0$ and $v_1$ can in practice be ignored---e.g., a {\sc not} gate can be implemented by a {\sc cmos} inverter in saturation; the input and output bit values 0 and 1 are encoded as different voltages $v_0$ and $v_1$, and the output voltage is essentially insensitive to small variations $\delta v \ll v_0, v_1$ in the input voltage. 

In the case of quantum computation, it is important to recognize that there is no unambiguous way to say which qubits of an entangled multi-qubit quantum state processed by the quantum computer are erroneous and which are not---e.g., consider the maximally entangled two-qubit state 
\begin{equation}
|\Phi_0\rangle = {1\over \sqrt{2}}(|0\rangle |0\rangle + |1\rangle |1\rangle) \; ; 
\end{equation}

\noindent if noise acts on $|\Phi_0\rangle$, there is no good way to say which of the two qubits is erroneous because for any single-qubit operator $E$, $(E\otimes I) |\Phi_0\rangle = (I\otimes E^T) |\Phi_0\rangle$, and whether the state has suffered an error cannot be determined by just observing the properties of any one of the two qubits in isolation. It is therefore helpful to avoid using a {\em semantic} language where the notion of an error depends on the quantum state on which errors act; instead, we adopt a {\em syntactic} language where the notion of an error is defined operationally independent of the actual quantum state. 
Instead of associating errors with individual qubits, we can associate errors with mutually orthogonal {\em subspaces} of the entire Hilbert space of a collection or a {\em block} of several encoded qubits; for a block comprising $n$ encoded qubits
\begin{equation}
\label{eq:discretize}
\mathcal{H}_n = \bigoplus_{s} \mathcal{H}_n^s \; ,
\end{equation}

\noindent where the superscript $s$, called the {\em syndrome}, is a label for the different subspaces $\mathcal{H}_n^s$ for the $n$-qubit Hilbert space $\mathcal{H}_n$. One of the subspaces, the {\em code space} $\mathcal{H}_n^0$, is the preferred one in the sense that quantum information is encoded in a quantum state that is supported in $\mathcal{H}_n^0$. Because of noise, the encoded quantum state will tend to escape from the code space toward other subspaces. But because all subspaces are mutually orthogonal, there is a generalized measurement, called a {\em syndrome  measurement}, which allows different subspaces to be distinguished unambiguously. By performing a syndrome measurement, we can then use the measurement outcome $\mu$ to deduce $s$ and therefore learn about the subspace on which the noisy encoded quantum state is supported. Since there is only a finite number of orthogonal subspaces, we can execute the computation that yields $s$ given $\mu$ digitally by processing the measurement outcome in a classical computer. If the result of this computation is a nontrivial syndrome value which indicates that the noisy encoded quantum state is supported in a subspace different than the code space, we can apply a {\em recovery} operation on the encoded qubits, which is conditioned on the result of the classical computation and which returns the encoded quantum state to the code space. Because classical digital computers are in practice extremely robust to noise, we usually assume that the classical processing of $\mu$ to obtain $s$ and the classical control of the quantum computer conditioned on $s$ can be implemented perfectly without faults. Of course, no matter how improbable, faults in the classical processing of the measurement outcomes can lead to a failure to apply the appropriate recovery operation, so that the accuracy of the quantum computer is ultimately limited by the accuracy of the on-the-side classical computer.

\section{Noisy quantum computers} \vspace{.3cm}

We have seen that discretizing the entire Hilbert space of the qubits processed by the quantum computer allows us to encode quantum information in a quantum state supported in the code space and to protect against noise that takes this encoded quantum state to other orthogonal subspaces. But for what types of noise processes is this method of encoding quantum information effective? Certainly, we have very little hope of protecting information against noise that acts collectively on many hardware components of the quantum computer and whose strength is not moderated as a function of the number of qubits it affects---e.g., we would be helpless if a power outage, an earthquake, or a high-energy cosmic ray\footnote[1]{High-energy cosmic rates are unlikely on earth, but they are a real concern for computers inside space shuttles.} were to hit our quantum computer affecting many qubits all at once.

If we exclude such malicious types of collective noise against which no error-correction method can be effective neither for quantum nor for classical computation, we are left with several other contributions to noise that need to be considered: First, there is noise due to imprecisions in the implementation of each elementary hardware operation---e.g., noise in the control parameters during the implementation of a unitary gate $U$ might result in realizing another unitary $U{+}\delta U$ instead, where $\delta U$ may be systematic or it may vary stochastically. Secondly, there are unwanted interactions among the qubits in the quantum computer---e.g., an electromagnetic coupling of nearby quantum-dot qubits which decays as a power of their relative distance. Thirdly, there are interactions between the qubits of the quantum computer and an {\em environment} representing external degrees of freedom which are not under our control---e.g., a coupling of integrated superconducting qubits to nuclear spins in the substrate. And finally, in settings where qubits are realized by selecting a two-dimensional subspace inside a multi-dimensional system, there is noise which couples the two-level qubit subspace to other levels of the same system---e.g., if a qubit is realized by using the ground-state hyperfine splittings of a trapped ion, noise can induce transitions between these hyperfine levels and other higher-energy levels of the ion. 

A useful classification of these noise processes concerns their spatial and temporal locality. Intuitively, we say that noise is spatially local or simply {\em local} if, during any time interval, it acts collectively only on qubits which are interacting in the ideal quantum circuit at the same time interval---i.e., if two-qubit gates are applied in parallel to several pairs of qubits during a specific time interval, local noise can act collectively on  qubits that belong to the same pair but not on those that belong to different pairs. Locality is a desirable property because it implies that noise cannot afflict global damage by causing the simultaneous failure of many hardware components. Although in general error correction fails for non-local noise, there are in fact some non-local noise processes for which effective error correction is possible: 
First, certain types of collective noise whose nature is known in advance can be suppressed effectively by using the techniques of decoherence-free subspaces and noiseless subsystems. 
And secondly, there are types of non-local noise which can be treated as if noise were local and for which fault-tolerance methods designed to protect against local noise are effective; we will discuss two such examples in Section \ref{sec:local-non-Markovian}. 

With regard to temporal locality, the question is how correlated is the noise that acts on different hardware components which are executed at different time intervals. We say that noise is temporally local or {\em Markovian} if the noisy evolution can be described by using a sequence of {\em superoperators} each taking the density matrix at the end of one time interval to the density matrix at the end of the following time interval.

When the quantum computer implements a quantum circuit of depth\footnote[1]{The {\em depth} of a quantum circuit is the maximum number of elementary operations applied on any qubit of the quantum computer (including the identity operation which is implicitly applied when a qubit is stored while operations are applied on other qubits).} $D$, we can discretize the total computation time $T$ in $D$ intervals $T_1, T_2, \dots, T_D$ each of duration $t_0$ equal to the time it takes to execute an elementary operation. The Markovian property then translates to the requirement that noise has a typical correlation time comparable to $t_0$. Since the interaction between the quantum computer and the environment is incoherent across different time intervals, we can trace over the state of the environmental degrees of freedom after each interval to obtain a reduced density matrix describing the state of the quantum computer. In this case, the noisy evolution is described as a mapping between the reduced density matrices at different intervals;
\begin{equation}
\rho_{j} = \mathcal{S}_{j} (\rho_{j-1}) \; ,
\end{equation} 
\noindent where $\rho_j$ is the reduced density matrix at the end of interval $T_j$, and $\mathcal{S}_{j}$ is a superoperator describing the evolution from the end of interval $T_{j-1}$ to the end of interval $T_j$.

On the other hand, if there are noise processes with typical correlation times longer than $t_0$, we cannot obtain an accurate description of the noisy evolution by tracing out the external degrees of freedom after every time interval. In this case, the information that the environment exchanges with the quantum computer could in principle be retained for long times so that we cannot simply describe the entire noisy evolution as a composition of superoperators. Because of this reason, the analysis of the effects of non-Markovian noise is more demanding than for simple Markovian noise and, as we shall see in the next section, our conclusions for the effectiveness of fault-tolerance methods against non-Markovian noise are generally weaker than for Markovian noise.

In this section, we will discuss several concrete examples of noise models that have been analyzed in the context of fault-tolerant quantum computation. 

\subsection{Setup} \vspace{.3cm}
\label{sec:noise-models}

We consider the noisy implementation of an ideal quantum circuit comprising $L$ elementary operations followed by the final qubit measurements whose outcome encodes the result of the computation. The ideal quantum circuit produces the quantum state 
\begin{equation}
\label{eq:ideal-circuit}
\rho^{\rm ideal} = \mathcal{O}_{L} \circ \dots \circ \mathcal{O}_{2} \circ \mathcal{O}_{1} \; ,
\end{equation}
where the superoperators ${\mathcal{O}}_j$ correspond to either a qubit preparation, a unitary gate, or an intermediate qubit measurement which conditions subsequent operations; the operation for the preparation of a qubit in the pure state $|\psi\rangle$ is 
\begin{equation}
\mathcal{P}_{|\psi\rangle} = |\psi\rangle \langle \psi| \; ,
\end{equation} 
\noindent the operation for a quantum gate $U$ applied on input $X$ is 
\begin{equation}
\mathcal{U}(X) = U X U ^\dagger \; ,
\end{equation}
\noindent and the operation for a projective measurement of an observable $\hat a$ applied on input $X$ with projector $M_a$ corresponding to measurement outcome $a$, 
\begin{equation}
\mathcal{M}_{\{a\}}(X) = \sum_a M_a\, X M_a \; . 
\end{equation}

For simplicity and since noise on the final measurements can be modeled by noise acting in the immediately preceding operations, we may assume that the final measurements are implemented ideally without faults. Finally, as we have noted, we will consider performing the processing of the outcomes of both the intermediate and the final measurements in a classical computer operating alongside the quantum computer, and we will assume there are practically no faults in this classical hardware.

\subsection{Local Markovian noise} \vspace{.3cm}
\label{sec:local-Markovian}

Our first example is noise which is both local and Markovian. The Markovian property implies that we can describe the noisy evolution as a composition of superoperators, and the locality property implies that the superoperator for each time interval can be expressed as a tensor product of superoperators, each superoperator corresponding to one of the different elementary operations which are implemented in parallel during that interval. 

We may express the superoperator describing the noisy implementation of each elementary operation as $\mathcal{N}_j \circ \mathcal{O}_j$, where $\mathcal{O}_j$ is the ideal superoperator and $\mathcal{N}_j$ is a superoperator describing deviations from the ideal due to noise---by definition, the support of $\mathcal{N}_j$ is contained in the support of $\mathcal{O}_j$ when noise is local\footnote[1]{The support of a superoperator $\mathcal{O}$ (or operator $O$) acting on density matrices (or quantum states respectively) defined on various subsystems is the tensor product of the Hilbert spaces of all the subsystems on which $\mathcal{O}$ (or $O$ respectively) acts nontrivially; in our case, the subsystems are the qubits of the quantum computer and any subsystems in the environment.}. Therefore, because of noise, instead of the ideal quantum state in eq.~(\ref{eq:ideal-circuit}), the quantum computer really prepares the state  
\begin{equation}
\label{eq:noisy-circuit}
\rho^{\rm noisy} = \mathcal{N}_L \circ \mathcal{O}_L \circ \dots \circ \mathcal{N}_2 \circ \mathcal{O}_2 \circ \mathcal{N}_1 \circ \mathcal{O}_1  \; ,
\end{equation}
\noindent where $\circ$ denotes composition.

Since all $\mathcal{N}_j$ would be trivial were there no noise, a natural measure for the noise is the distance between $\mathcal{N}_j$ and the identity superoperator $\mathcal{I}$, and we can define the noise {\em strength}
\begin{equation}
\label{eq:noise-strength}
\varepsilon = \max\limits_j || \mathcal{N}_j - \mathcal{I} ||_\diamond \; ,
\end{equation}
\noindent where $||\cdot||_{\diamond}$ is a suitable superoperator norm, the {\em diamond} norm
\footnote[2]{If the superoperator ${\cal E}$ has an $n$-qubit input, and ${\cal I}_n$ is the identity superoperator on $n$ qubits, then $||{\cal E}||_{\diamond} =  \max ||({\cal E} \otimes I_n)(X)||_1$, where we maximize over all $X$ such that $||X||_1{=}{\rm Tr}\sqrt{X^\dagger X}{=}1$. Note that we could optimize our estimate for the noise strength by taking $\mathcal{I}$ to be proportional to the identity superoperator with a proportionality constant of magnitude between 0 and 1; but here we will not discuss this generalization.}.
If we now write $\mathcal{N}_j = \mathcal{I} + \mathcal{F}_j$ for some fault operator, or simply {\em fault}, $\mathcal{F}_j$ and substitute in eq.~(\ref{eq:noisy-circuit}), we find
\begin{equation}
\rho^{\rm noisy} = (\mathcal{I} + \mathcal{F}_L) \circ \mathcal{O}_L \circ \dots \circ (\mathcal{I} + \mathcal{F}_2) \circ \mathcal{O}_2 \circ (\mathcal{I} + \mathcal{F}_1 ) \circ \mathcal{O}_1 \; .
\end{equation}
\noindent By opening all parentheses, we obtain a sum of terms corresponding to different {\em fault paths}; in each fault path, faults $\mathcal{F}_j$ have occurred in a specific subset of the $L$ elementary operations, while the identity superoperators are applied on all remaining operations. In particular, we can write 
\begin{equation}
\label{eq:fault-path-expansion}
\rho^{\rm noisy} = \rho^{\rm ideal} + \zeta^{\rm faulty} \; ,
\end{equation}
\noindent where $\rho^{\rm ideal}$ corresponds to the unique fault path where identity superoperators are applied everywhere and $\zeta^{\rm faulty}$ contains all other fault paths for which there is at least one insertion of a fault---we use $\zeta$ instead of $\rho$ in $\zeta^{\rm faulty}$ to emphasize that it is not a density matrix but rather the difference of two density matrices. 

The fault path expansion is helpful for understanding how accurate is the noisy circuit. More precisely, we would like to know what is the distance $\delta$ between the probability distribution $\{ q_{\mu}^{\rm noisy}\}$ for the outcomes $\{ \mu \}$ of the final measurements on the noisy circuit and the distribution $\{q_{\mu}^{\rm ideal}\}$ if these measurements were applied on the ideal circuit instead. We can express $\delta$ in terms of the 1-norm (or Kolmogorov distance) between the two probability distributions, 
\begin{equation}
\delta = \sum_\mu |q_{\mu}^{\rm noisy} - q_{\mu}^{\rm ideal}| = \sum_\mu | {\rm Tr}\left(M_\mu \left( \rho^{\rm noisy} - \rho^{\rm ideal} \right) \right)|\; ,
\end{equation}
\noindent where the projectors $M_\mu$ define the final measurements\footnote[1]{The projectors $M_\mu$ are non-negative (i.e., $\langle u | M_\mu |u\rangle\geq 0$ for any $|u\rangle$), and they are normalized so that $\sum_\mu M_\mu =I$.}. But next, $\delta$ can be related to the trace norm between $\rho^{\rm noisy}$ and $\rho^{\rm ideal}$ 
; i.e., 
\begin{equation}
\label{eq:base-accuracy}
\delta = \sum_\mu | \sum_\nu v_\nu \langle v_\nu | M_\mu |v_\nu\rangle | \leq \sum_\mu \sum_\nu |v_\nu| |\langle v_\nu|M_\nu|v_\nu \rangle| \leq \sum_\nu |v_\nu| = || \zeta^{\rm faulty} ||_1 \; , 
\end{equation} 
\noindent where $|v_\nu\rangle$ are the eigenvectors of $\zeta^{\rm faulty} = \rho^{\rm noisy} - \rho^{\rm ideal}$ with corresponding eigenvalues $v_\nu$. 

Now, how do we upper bound the norm of $\zeta^{\rm faulty}$? If we let $C$ denote the set of all $L$ elementary operations in the quantum circuit, 
\begin{equation}
\label{eq:fault-paths}
\zeta^{\rm faulty} = \sum\limits_{r=1}^{L} (-1)^{r-1} \sum\limits_{C_r \subseteq C} \zeta(C_r) \; ,
\end{equation} 
\noindent where the second sum is over all subsets $C_r$ of $C$ with cardinality $r$, and $\zeta(C_r)$ denotes a {\em sum} of all the fault paths with faults applied on all operations in the set $C_r$. Eq.~(\ref{eq:fault-paths}) can be derived from the {\em inclusion-exclusion} trick of combinatorics: Since $\zeta^{\rm faulty}$ is the sum of all the fault paths with at least one fault, the sum of all $\zeta(C_1)$ counts correctly all the fault paths with exactly one fault but overcounts the fault paths with at least two faults; to amend the overcounting, we subtract the sum of all $\zeta(C_2)$ which corrects the overcounting of all the fault paths with exactly two faults but introduces an undercounting of the fault paths with at least three faults; and so on.

For each specific set $C_r$, $\zeta(C_r)$ is nothing by the composition of the ideal superoperators $\mathcal{O}_j$ interspersed with faults $\mathcal{F}_j$ applied on all operations in $C_r$ and the full noise superoperators $\mathcal{N}_j$ applied on all the remaining operations. Of course, since superoperators have unity norm, $||\mathcal{O}_j||_\diamond = ||\mathcal{N}_j ||_\diamond =1 $ and thus
\begin{equation}
\label{eq:local-noise-definition}
||\zeta(C_r)||_1 \leq \varepsilon^r \; .
\end{equation}
\noindent By using the triangle inequality, and since there are ${L\choose r}$ distinct subsets $C_r$ of $C$, eqs.~(\ref{eq:base-accuracy}) and (\ref{eq:fault-paths}) now imply that
\begin{equation}
\label{eq:markovian-accuracy}
\delta \leq \sum\limits_{r=1}^L {L \choose r} \varepsilon^r \leq L\varepsilon \left( 1 + {1\over L^2} {L\choose 2} + {1\over L^3} {L\choose 3} + \cdots + {1\over L^L} {L\choose L} \right) \leq (e{-}1)L\varepsilon \; , 
\end{equation}
\noindent where in the last two steps we assumed that $\varepsilon\leq 1/L$. 

Our derivation of eq.~(\ref{eq:markovian-accuracy}) via eq.~(\ref{eq:fault-paths}) was made in order to introduce a simple application of the inclusion-exclusion trick that is also being used later in this chapter. An improved upper bound on $\delta$ can in fact be derived without the assumption $\varepsilon\leq 1/L$ by simply noting that we can group fault paths depending on their {\em earliest} faulty operation;  
\begin{equation}
\label{eq:fault-paths2}
\zeta^{\rm faulty} = \sum\limits_{r=1}^{L} \zeta(\mathcal{O}_r) \; ,
\end{equation} 
\noindent where $\zeta(\mathcal{O}_r)$ is the composition of the ideal superoperators $\mathcal{O}_j$ interspersed with the identity superoperators applied on the operations 1 to $r{-}1$, a fault $\mathcal{F}_r$ applied on the $r$-th operation, and the full noise superoperators $\mathcal{N}_j$ applied on the operations $r{+}1$ to $L$. Since superoperators have unity norm, $||\zeta(\mathcal{O}_r)||_\diamond \leq \varepsilon $ and thus eq.~(\ref{eq:fault-paths2}) implies 

\begin{equation}
\label{eq:markovian-accuracy2}
\delta \leq L\varepsilon \; .
\end{equation}
We conclude that, for a constant error strength $\varepsilon$, the accuracy $1\,{-}\,\delta$ of the noisy quantum circuit decreases at most {\em linearly} with the circuit size $L$, in accordance to what is expected for a discrete model of computation\footnote[1]{For discrete models of computation, to achieve a constant accuracy $1\,{-}\,\delta$, the number of bits of precision required to specify the physical parameters associated with each elementary operation to within $\varepsilon$---e.g., the amplitude and timing of a voltage pulse used to control a {\sc CMOS} gate---grows logarithmically with the size $L$ of the computation. (With the bits of precision growing logarithmically with $L$, $\varepsilon$ decreases polynomially with $L$.) In contrast, for analog models, the number of bits of precision grows polynomially or even exponentially with $L$.}. Of course, as we shall discuss in the following sections, the goal of implementing the quantum computation by using fault tolerance methods is to replace $\varepsilon$ in eq.~(\ref{eq:markovian-accuracy}) by a smaller---in fact, an arbitrarily small---{\em effective} noise strength, thus making the accuracy of the noisy circuit approach as close to unity as desired. 

\subsubsection{Assessment and examples}

At this point, we can step back to note the two essential assumptions that allowed us to derive eq.~(\ref{eq:markovian-accuracy2}): First, we assumed that the superoperators describing the noisy evolution can be expanded perturbatively as a sum over fault paths. Secondly, we assumed that fault paths with many faults are exponentially suppressed in the sense of eq.~(\ref{eq:local-noise-definition}). In fact, we may view eq.~(\ref{eq:local-noise-definition}) as the {\em defining} property of local Markovian noise, even if the superoperators that describe the noisy evolution are not strictly local. Thus, we generally say that 

\vspace{.3cm}
{\bf Definition} (Local Markovian noise). {\em Noise is local and Markovian if the noisy evolution can be expanded as a sum over fault paths, where faults are described as (differences of) {\em superoperators} and the norm of the sum of all the fault paths with faults in any $r$ specific elementary operations is upper bounded by $\varepsilon^r$ for some constant noise strength $\varepsilon$.}
\vspace{.3cm}

This relaxed definition has the advantage that it can describe correlated noise both in space and in time: Subject to the constraint that fault paths must satisfy eq.~(\ref{eq:local-noise-definition}), the fault operators comprising each fault path are otherwise unconstrained; in particular, the various fault operators can be controlled by an adversary who may chose to act collectively on all the faulty operations any way she pleases. 

The local Markovian noise model captures several noise processes of interest; below, we discuss three simple but important examples. 
 
\vskip 0.0cm \HRule \vskip 0.1cm
\noindent {\bf Control noise.} A common source of noise is due to imprecision in the control parameters during the implementation of each elementary operation---e.g., noise in the timing or the intensity of external magnetic fields used to manipulate the state of a superconducting qubit. 

In the simplest case, consider the implementation of a single-qubit gate corresponding to a rotation by an angle $2\theta$ around the z direction; this operation is described by the superoperator
\begin{equation}
\mathcal{R}^{\rm z}_\theta (X) = e^{i\theta\sigma_{\rm z}} X e^{-i\theta\sigma_{\rm z}} \; .
\end{equation} 

Because of imprecisions in the control parameters, a rotation by a different angle $2\theta' =2(\theta{+}\delta\theta)$ for some small fixed deviation $\delta\theta$ may be implemented instead. We can express the noisy superoperator as $\mathcal{R}^{\rm z}_{\theta'} = \mathcal{N}_{\rm ctrl} \circ \mathcal{R}^{\rm z}_\theta$, where $\mathcal{N}_{\rm ctrl} = \mathcal{I} + \mathcal{F}_{\rm ctrl}$ and 
\begin{equation}
\mathcal{F}_{\rm ctrl} (X) = i\, \delta\theta\, ( \sigma_{\rm z}\, X - X\, \sigma_{\rm z} ) + O(\delta\theta^2) \; ,
\end{equation}
\noindent so that control noise satisfies eq.~(\ref{eq:local-noise-definition}) with $\varepsilon = O(\delta\theta)$. 

A similar conclusion also holds if the deviation angle is not fixed but varies stochastically, and also for control errors in multi-qubit gates, preparations or measurements.
\vskip 0.1cm \HRule \vskip 0.1cm
\noindent {\bf Relaxation.} Another common source of noise is due to thermal relaxation---e.g., in systems where $|0\rangle$ and $|1\rangle$ are encoded in different energy eigenlevels, the state $|1\rangle$ may spontaneously relax to the lower-energy state $|0\rangle$. To first approximation, relaxation can be expected to act independently on each qubit during the execution of a quantum computation; then for each qubit and during each time interval, relaxation with a characteristic time scale $T_1$ can be modeled by the {\em amplitude damping} superoperator $\mathcal{N}_{\rm relax} (X) = M_0 X M_0^\dagger + M_1 X M_1^\dagger$, where 
\begin{equation}
M_0 = {1 + \sqrt{1-\gamma} \over 2} \; I + {1-\sqrt{1-\gamma}\over 2} \; \sigma_{\rm z} \; \; , \; M_1 = {\sqrt{\gamma}\over 2} \sigma_{\rm x} (1 - \sigma_{\rm z}) \; ,
\end{equation}
\noindent and $\gamma = 1-e^{-t_0/T_1}$. We can write $\mathcal{N}_{\rm relax} = \mathcal{I} + \mathcal{F}_{\rm relax}$ where 
\begin{equation}
\mathcal{F}_{\rm relax} (X) = {\gamma \over 4} \sigma_{\rm z} X M_0^\dagger + {\gamma \over 4} M_0 X \sigma_{\rm z} + M_1 X M_1^\dagger + O(\gamma^2) \; ,
\end{equation}
\noindent so that relaxation noise satisfies eq.~(\ref{eq:local-noise-definition}) with $\varepsilon = O(\gamma)$.
\vskip 0.1cm \HRule \vskip 0.1cm
\noindent {\bf Probabilistic noise.} In many cases noise can be modeled as a random processes---e.g., shot noise in the laser fields used to control trapped ionic qubits. Ignoring the details of the underlying random process, and letting $p$ denote the probability of a fault during the implementation of each elementary operation---if $r$ is the fault rate, $p = r t_0$---, noise can be modeled by the superoperator 
\begin{equation}
\mathcal{N}_{\rm rand} (X) = (1-p) X + p E X E^\dagger \; ,
\end{equation} 
\noindent where $E$ is an arbitrary operator acting on the support of the ideal operation (subject to the constraint $E^\dagger E = I$ required for $\mathcal{N}_{\rm rand}$ to be trace preserving). 

We can write $\mathcal{N}_{\rm rand} = \mathcal{I} + \mathcal{F}_{\rm rand}$ where 
\begin{equation}
\mathcal{F}_{\rm rand} (X) = -p X + p E X E^\dagger \; ,
\end{equation}
\noindent so that probabilistic noise satisfies eq.~(\ref{eq:local-noise-definition}) with $\varepsilon = O(p)$. 
\vskip 0.1cm \HRule \vskip 0.3cm

\subsection{Local non-Markovian noise} \vspace{.3cm}
\label{sec:local-non-Markovian}

Our local Markovian noise model is powerful enough to capture several important noise processes such as systematic control errors or thermal relaxation. However, requiring that fault paths are associated with superoperators is rather limiting because it constrains the possible noise correlations between different fault paths; while the fault operators in any specific fault path may be arbitrarily correlated, there can only be classical but no quantum correlations---i.e., no quantum interference---between the fault operators in different fault paths. 

To go beyond Markovian noise, we can no longer trace over the environmental degrees of freedom to obtain a superoperator description of the noisy evolution. Now, our description will need to include explicitly the quantum state of the environment and its joint unitary evolution with the qubits of the quantum computer during the course of the entire quantum computation---of course, while we assume that we have control over all the qubits of our quantum computer, the environmental degrees of freedom are inaccessible and in many cases their precise nature is unknown. 

\begin{figure}[t]
\begin{tabular}{c}
\put(-5.8,0){\includegraphics[width=13cm,keepaspectratio]{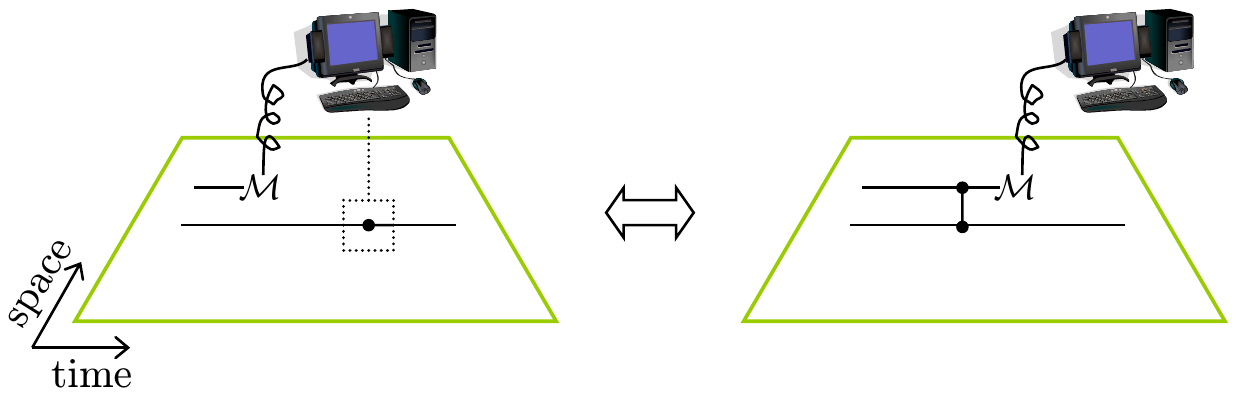}}
\vspace{-6.3cm}
\end{tabular}
\caption{\label{fig:meas-replace} On the left, a unitary gate $U$ in a quantum circuit is conditioned on the outcome of a preceding single-qubit measurement along the orthonormal basis $\{|\tilde 0\rangle, |\tilde 1\rangle \}$. On the right, a mathematically equivalent circuit where the two qubits interact via a unitary gate $\tilde U$ followed by a single-qubit measurement along the basis $\{|0\rangle, |1\rangle \}$; here, $\tilde U = \Lambda(U) (V \otimes I) $ where the single-qubit unitary $V$ acts on the measured qubit as $V|\tilde 0\rangle=|0\rangle$,  $V|\tilde 1\rangle=|1\rangle$, and the two-qubit $\Lambda(U)$ applies $U$ on the second qubit conditioned on the state of the measured qubit being $|1\rangle$.}
\end{figure}

We assume that the qubits of the quantum computer can be initialized in a pure state $|\psi_0\rangle_{QC}$---e.g., we can prepare all qubits in their lowest-energy eigenstate (at least, come very close to it) by cooling---so that the state at the beginning of the quantum computation, including the environment, is $|\psi_0\rangle_{QC} \otimes |\phi_0\rangle_E$ for some unspecified pure state\footnote[1]{We can always obtain a representation of the environment in terms of a pure state since any mixed state can be purified by introducing an auxiliary Hilbert space.} $|\phi_0\rangle_E$. 
If there were no noise, implementing the ideal quantum circuit would then correspond to implementing a sequence of unitary operators $U_j$ producing the final state 
\begin{equation}
\label{eq:ideal-circuit-pure}
|\psi\rangle_{QC}^{\rm ideal} = \left( U_{L} \cdots U_{2} \cdot U_{1} \right) |\psi_0\rangle_{QC} \; ,
\end{equation}
\noindent on which state we finally apply measurements that give the result of the computation. (If there are unitary gates that are conditioned on the outcome of a preceding measurement---cf. fig.~\ref{fig:circuits}---, we can mathematically replace them in our analysis by different unitary gates which are followed by measurements as in fig.~\ref{fig:meas-replace}.)

We may describe the noisy implementation of each elementary unitary operator $U_j$ as $N_j \cdot U_j$ where $N_j$ is a unitary operator on the quantum computer {\em and} the environment and it describes deviations from the ideal due to noise---by definition, the support of $N_j$ is contained in the union of the support of $U_j$ and the environment when noise is local. Since all $N_j$ would be trivial were there no noise, we can now define the noise strength as    
\begin{equation}
\label{eq:noise-strength-pure}
\varepsilon = \max\limits_j \min_{I_{QC}} || N_j - I_{QC} ||_{\infty} \; ,
\end{equation}
\noindent where 
we vary over all unitary $I_{QC}$ that act trivially on the quantum computer and in an arbitrary way on the environment.

By expanding each $N_j$ as the sum of the $I_{QC}$ that minimizes the norm in eq.~(\ref{eq:noise-strength-pure}) and a {\em fault}, $N_j = I_{QC} + F_j$, we can substitute in eq.~(\ref{eq:ideal-circuit-pure}) to find that the joint state of the quantum computer and the environment prior to the final measurements is
\begin{equation}
|\psi\rangle_{QCE}^{\rm noisy} = (I_{QC} + F_L) \cdot U_L \cdots (I_{QC} + F_2) \cdot U_2 \cdot (I_{QC} + F_1 ) \cdot U_1 (|\psi_0\rangle_{QC} \otimes |\phi_0\rangle_E ) \; ,
\end{equation}
\noindent where it is understood that the $U_j$ act trivially on the environment. We may now open all the parentheses to obtain a fault-path expansion just like we did in the previous section; here, each fault path identifies a specific subset of the $L$ elementary operations where faults $F_j$ have occurred, where the $F_j$ are differences of unitary operators instead of differences of superoperators as in the case of Markovian noise. We can then write
\begin{equation}
\label{eq:fault-path-expansion-pure}
|\psi\rangle^{\rm noisy} = |\psi\rangle^{\rm ideal} + |\vartheta\rangle^{\rm faulty} \; ,
\end{equation}
\noindent where for succinctness we have dropped the $QCE$ subscripts but it is understood that all states are supported in the Hilbert space of the qubits of the quantum computer and also the environment. Here, $|\psi\rangle^{\rm ideal}$ corresponds to the unique fault path where operators $I_{QC}$ are applied everywhere and $|\vartheta\rangle^{\rm faulty}$ contains all other fault paths for which there is at least one insertion of a fault---we use $\vartheta$ instead of $\psi$ in $|\vartheta\rangle^{\rm faulty}$ to emphasize that it is not normalized but it is rather the difference of two normalized pure states. 

How can we estimate the accuracy of the noisy quantum circuit in the presence of local non-Markovian noise? From eq.~(\ref{eq:markovian-accuracy}), it suffices to evaluate the trace norm of the difference between the noisy and the ideal density matrices; since the final states are pure,
\begin{equation}
\label{eq:base-accuracy-non-Markovian}
\delta = || |\psi\rangle \langle \psi|^{\rm noisy} - |\psi\rangle \langle \psi|^{\rm ideal} ||_1 \leq 2 || |\vartheta\rangle^{\rm faulty} ||  \; , \vspace{.05cm}
\end{equation} 
\noindent where we used that $|||n\rangle \langle n|{-}|i\rangle \langle i| ||_1 = 2 \sqrt{1-|\langle n| i \rangle|^2} \leq 2 || |n\rangle{-}|i\rangle ||$ for any normalized pure states $|n\rangle$ and $|i\rangle$. It remains to obtain an upper bound on the norm  of $|\vartheta\rangle^{\rm faulty}$.

In analogy to eq.~(\ref{eq:fault-paths}),
\begin{equation}
|\vartheta\rangle^{\rm faulty} = \sum\limits_{r=1}^{L} (-1)^{r-1} \sum\limits_{C_r \subseteq C} |\vartheta(C_r)\rangle \; ,
\end{equation} 
\noindent where $|\vartheta(C_r)\rangle$ denotes a {\em sum} of all the fault paths with faults applied on all $r$ operations in the set $C_r$; since $|\vartheta(C_r)\rangle$ is obtained by applying unitary operators everywhere except at the faulty operations, 
\begin{equation}
\label{eq:local-noise-definition-pure}
|||\vartheta(C_r)\rangle|| \leq \varepsilon^r \; .
\end{equation}

\noindent Alternatively, in analogy to eq.~(\ref{eq:fault-paths2}), we may group the fault paths depending on the earliest faulty operation;
\begin{equation}
|\vartheta\rangle^{\rm faulty} = \sum\limits_{r=1}^{L} |\vartheta(U_r)\rangle \; ,
\end{equation} 
\noindent where $|| |\vartheta(U_r)\rangle || \leq \varepsilon$. We conclude that
\begin{equation}
\label{eq:non-markovian-accuracy}
\delta \leq 2 L \varepsilon \; .
\end{equation} 
\noindent so that we obtain for local non-Markovian noise a similar result as for local Markovian noise---in both cases, the accuracy $1\,{-}\,\delta$ of the noisy quantum circuit decreases at most linearly with the circuit size $L$. This illustrates that fully coherent noise, where the environment can store and process quantum information allowing different fault paths to interfere quantum mechanically, does not alter our view of quantum computation as a discrete model of computation similar to the model of modern digital computers (or, more abstractly, classical Turing machines). As we shall discuss in the following sections, methods of quantum fault tolerance can replace $\varepsilon$ in eq.~(\ref{eq:non-markovian-accuracy}) by an arbitrarily small effective noise strength showing that, just like Markovian noise, non-Markovian noise is not in principle an obstacle to large-scale quantum computation. 

\subsubsection{Assessment and examples}

We note that eq.~(\ref{eq:non-markovian-accuracy}) was derived based on two essential assumptions: First, we assumed that the final noisy quantum state can be expanded perturbatively as a sum over fault paths. Secondly, we assumed that fault paths with many faults are suppressed in the sense of eq.~(\ref{eq:local-noise-definition-pure}). We can in fact {\em define}  local non-Markovian noise in terms of these two assumptions, even if the noisy unitary evolution is not strictly local. We then generally say that

\vspace{.3cm}
{\bf Definition} (Local non-Markovian noise). {\em Noise is local and non-Markovian if the noisy evolution can be expanded as a sum over fault paths, where faults are described as (differences of) {\em unitaries} acting between the quantum computer and the environment and the norm of the sum of all the fault paths with faults in any $r$ specific elementary operations is upper bounded by $\varepsilon^r$ for some constant noise strength $\varepsilon$.}
\vspace{.3cm}

This definition is very similar to our definition of local Markovian noise in the previous section: In both cases, the noisy evolution is expanded as a sum over fault paths and also noise is {\em weak} in the sense that, as the total number of faults in a fault path increases, the fault path norm is suppressed exponentially. In addition, in both cases noise can be correlated both in space and in time since we place no restrictions on the form of the fault operators that appear in each fault path, which are allowed to be arbitrarily and even adversarially correlated. 

The important distinction between the two cases is that while, for local Markovian noise, different fault paths do not interfere, for local non-Markovian noise, we make the worst-case assumption that all the fault paths {\em do} interfere coherently and the environment is {\em not} traced over until the end of the quantum computation. Therefore, while for local Markovian noise the strength $\varepsilon$ can be viewed as a probability (the probability for the occurrence of a single fault), for local non-Markovian noise the strength $\varepsilon$ corresponds in essence to a quantum {\em amplitude} (the amplitude for a term with a single fault in the final quantum state). If we were to use distinct symbols for the two cases, $\varepsilon$ and $\varepsilon'$ respectively for Markovian and non-Markovian noise then, since probabilities are squares of amplitudes, we expect $\varepsilon \sim \left( \varepsilon'\right)^2$; thus, requiring that the fault amplitude $\varepsilon'$ is small (say, less than $1.0\times 10^{-3}$) implies that the fault probability $\varepsilon$ is even smaller (in this case less than $1.0\times 10^{-6}$). 

The local non-Markovian noise model describes several noise processes for which the environment interacts {\em coherently} with the quantum computer over long time scales; below, we discuss four examples of noise processes that give rise to local non-Markovian noise.

\vskip 0.0cm \HRule \vskip 0.1cm
\noindent {\bf Local Hamiltonian noise.} The ideal unitary evolution in eq.~(\ref{eq:ideal-circuit-pure}) is generated by a time-dependent Hamiltonian
\begin{equation}
H_{QC} = \sum\limits_j H_{QC}^j \; , \; {\rm such}\,\, {\rm that}\,\, U_j = \exp\left(-it_0 H_{QC}^j \right) \; .
\end{equation}
\noindent The Hamiltonian that describes the noisy evolution of the quantum computer and the environment then has the general form
\begin{equation}
\label{eq:hamiltonian-evolution}
H = H_{QC} + H_E + H_{QCE} \; ,
\end{equation}
\noindent where $H_E$ generates the evolution of the environmental degrees of freedom, and $H_{QCE}$ describes the interaction of the quantum computer and the environment that introduces noise. If noise is local, $H_{QCE}$ has the same locality as $H_{QC}$, i.e.,
\begin{equation}
\label{eq:local-interaction}
H_{QCE} = \sum\limits_j H_{QCE}^j \; ,
\end{equation}
\noindent where the support of $H_{QCE}^j$ is contained in the union of the support of $H_{QC}^j$ {\em and} the environment.

We can study the noisy evolution generated by $H$ during a time interval of duration $t_0$ perturbatively; if we  divide this interval into $N$ micro-intervals each of duration $\Delta t_0 = t_0 / N$, then 
\begin{equation}
\label{eq:noisy-evolution-trotter}
U_{QCE}(t_0,0) = \lim\limits_{N\rightarrow\infty} \prod\limits_{n=1}^N U^n_{QC} U^n_{QCE} U^n_E \; ,
\end{equation}
\noindent where $U^n_{QC}$, $U^n_{QCE}$, and $U^n_{E}$ denote the evolution during the $n$-th micro-interval according to $H_{QC}$, $H_{QCE}$, and $H_E$ respectively. After expanding\footnote[1]{Since we take the limit $N{\rightarrow}\infty$, we have $\Delta t_0{\rightarrow}0$ and we may keep only the linear term in the expansion.}
\begin{equation}
U^n_{QCE} \approx \prod\limits_j \left( I - i\Delta t_0 H^j_{QCE} \right) \; ,
\end{equation}
\noindent and substituting in eq.~(\ref{eq:noisy-evolution-trotter}), we can open the parentheses to obtain a perturbative fault-path expansion. The noisy implementation of the ideal unitary $U_j$ then takes the form
\begin{equation}
U_j^{\rm noisy} = \left(I_{QC}+ F_j \right) U_j \; ;
\end{equation}
\noindent here, $I_{QC}$ denotes a sum of all the fault paths where, in every micro-interval, we insert either the identity or a {\em micro-fault} $f_{j'}$ acting on an operation with label $j'{\not = j}$ where
\begin{equation}
f_{j'} = - i\Delta t_0 H^{j'}_{QCE} \; .
\end{equation} 
\noindent It follows that $F_j$ includes all remaining fault paths where a micro-fault $f_j$ acting on the operation with label $j$ is inserted in at least one micro-interval.

We can express $F_j$ as a sum of terms labeled by the micro-interval where the {\em earliest} micro-fault $f_j$ is applied on the operation with label $j$; if we denote by $U_{QCE}(\Delta t_0^j,\Delta t_0^i)$ the entire evolution generated by $H_{QCE}$ between the $i$-th and $j$-th micro-intervals with $i<j$, then
\begin{equation}
\label{eq:fault-path-expansion-local-Hamiltonian-noise}
F_j = \lim\limits_{N\rightarrow\infty} \sum\limits_{r=1}^N \left( U_{QCE}(\Delta t_0^N,\Delta t_0^{r+1}) \prod\limits_{q=1}^{r} \left( U^{q}_{QC} f_j^{\delta_{r,q}} \prod\limits_{j' \not = j} \left( I +f_{j'} \right) U^{q}_E \right)  \right) U_j^\dagger \; .
\end{equation}
Now, since the operator norm is unitarily invariant, each term in the sum in eq.~(\ref{eq:fault-path-expansion-local-Hamiltonian-noise}) has norm $||f_j||_{\infty} = \Delta t_0||H^j_{QCE}||_{\infty}$, and so $||F_j||_{\infty} \leq t_0 ||H^j_{QCE}||_{\infty}$. 

In fact, we can perform a similar perturbative expansion to analyze faults in any specific subset of the $L$ elementary operations in the quantum circuit. It follows that local Hamiltonian noise satisfies eq.~(\ref{eq:local-noise-definition-pure}) with
\begin{equation}
\label{eq:local-Hamiltonian-noise-strength}
\varepsilon = t_0 \cdot \max\limits_j ||H^j_{QCE}||_{\infty} \; .
\end{equation}
\noindent As expected, the noise strength $\varepsilon$ depends on the strength of the interaction term $H^j_{QCE}$ between the quantum computer and the environment and also the time during which is interaction is acting. On the contrary, we observe that $\varepsilon$ does {\em not} depend on the strength of the term $H_E$ which describes the internal evolution of the environment. Moreover, we note that while $H^j_{QCE}$ is assumed to act locally on the quantum computer in the sense of eq.~(\ref{eq:local-interaction}), the derivation of eq.~(\ref{eq:local-Hamiltonian-noise-strength}) did not rely on making any assumptions about $H_E$ which is completely arbitrary.   


\vskip 0.1cm \HRule \vskip 0.1cm
\noindent {\bf Long-range static noise.} In certain systems, noise can arise due to static---i.e., time-independent---interactions among pairs of qubits of the quantum computer, where these interactions do not depend on the ideal circuit that is being implemented. Such non-local noise can be modeled by the Hamiltonian in eq.~(\ref{eq:hamiltonian-evolution}) with  
\begin{equation}
H_{QCE} = \sum\limits_{(j,k)} H_{(j,k)} \; ,
\end{equation} 
\noindent where $H_{(j,k)}$ is supported on qubits $j,k$ and the environment and we sum all unordered pairs $(j,k)$ of qubits.

We can perform a similar perturbative expansion as in the case of local Hamiltonian noise, except that now the two qubits in the support of any micro-fault may not be directly interacting via a unitary gate. Despite this difference which necessitates a more complicated combinatorial analysis (see the references), it can be shown that long-range static noise satisfies eq.~(\ref{eq:local-noise-definition-pure}) with
\begin{equation}
\varepsilon = \sqrt{c\,t_0\,\max\limits_{j} \sum\limits_k ||H_{(j,k)}||_{\infty} } \; , 
\end{equation}
\noindent where $c=2e$ provided $\varepsilon^2\leq e$ and it is understood that, if $H_{QCE}$ is time dependent, the maximum is also taken over all times.
%
\vskip 0.1cm \HRule \vskip 0.1cm
\noindent {\bf Gaussian noise.} In a variety of physical setting where the qubits of the quantum computer are coupled to a large number of environmental degrees of freedom, the environment can be well approximated as a collection of uncoupled harmonic oscillators obeying Gaussian statistics; the Hamiltonian of the environment is
\begin{equation}
H_E = \sum\limits_k \omega_k a_k^\dagger a_k \; ,
\end{equation}
\noindent where $a_k$ are bosonic annihilation operators satisfying $[a_k, a_{k'}^\dagger] = \delta_{kk'}$. In this {\em spin-boson model} of the noise, the interaction between the quantum computer and the environment is described by a coupling of each qubit to a linear combination of oscillator amplitudes  
\begin{equation}
H_{QCE} = \sum\limits_{x,m} \sigma_m(x)\otimes \tilde\phi_m(x,t) 
\end{equation}
\noindent with
\begin{equation}
\phi_m(x,t) = e^{itH_E} \tilde\phi_m(x,t) e^{-itH_E} = \sum\limits_k \left( g_{k,m}(x,t)a_k e^{-it\omega_k} + g^*_{k,m}(x,t)a_k^\dagger e^{it\omega_k} \right) \; , 
\end{equation}
\noindent where $x$ labels a qubit's position and $\sigma_m(x)$ with $m \in \{ x, y, z\}$ are the three Pauli operators on the qubit with label $x$. 

The statistics of the environment amplitudes $\phi_m(x,t)$ are Gaussian in the sense that the $n$-point correlation functions with respect to the environment state $|\phi_0\rangle_E$ vanish for $n$ odd, while for $n$ even they obey Wick's theorem:
\begin{equation}
\langle \phi_{m_1}(x_1,t_1) \cdots \phi_{m_n}(x_n,t_n) \rangle = \sum\limits_{(i_1,i_2), \cdots, (i_{n-1},i_n)} \Delta(i_1,i_2) \cdots \Delta(i_{n-1},i_n) \; , 
\end{equation}
\noindent where $\Delta(p,q) = \langle \phi_{m_{p}}(x_{p},t_{p}) \phi_{m_{q}}(x_{q},t_{q})\rangle$, and we sum all ways of dividing the label 1 to $n$ into $n/2$ unordered pairs. By performing a perturbative analysis similar to the case of long-range static noise (see the references), it can be shown that Gaussian non-Markovian noise satisfies eq.~(\ref{eq:local-noise-definition-pure}) with
\begin{equation}
\varepsilon = \sqrt{c \max\limits_{j} \int_{(x_1,t_1)\in U_j} \int_{(x_2,t_2)\in \cup_l U_l} \sum\limits_{m_1,m_2} \left| \Delta(1,2) \right| } \; , 
\end{equation}
\noindent where the first integral denotes an integration over the qubits in the support of the unitary $U_j$ and the time interval during which this gate is implemented, and the second integral denotes an integration over all the qubits of the quantum computer and the total duration of the quantum computation. 
%
\vskip 0.0cm \HRule \vskip 0.1cm
\noindent {\bf Local leakage noise.} The qubits of the quantum computer are in practice always realized as two-dimensional subspaces inside a multi-dimensional system; the Hilbert space $\mathcal{H}_{QC}$ of the quantum computer then has a natural extension to
\begin{equation}
\mathcal{H}_{QC}^{\rm ext} = \mathcal{H}_{QC} \oplus \mathcal{H}_{QC}^\perp \; ,
\end{equation}   
\noindent where the {\em leakage} space $\mathcal{H}_{QC}^\perp$ includes all states outside the two-dimensional qubit subspaces---in most settings, $\mathcal{H}_{QC}^\perp$ is a tensor product over leakage spaces corresponding to each qubit.   

Now, the Hamiltonian that describes the noisy evolution of the quantum computer and the environment has the same general form in eq.(\ref{eq:hamiltonian-evolution}), 
\begin{equation}
H = H_{QC} + H_{QC}^\perp + H_E + H_{QCE}^{\rm ext} \; ,
\end{equation}
where $H_{QC}^\perp$ generates the evolution in the leakage space, and $H_{QCE}^{\rm ext}$ describes the interaction between the extended space of the quantum computer (the qubits and their leakage spaces) and the environment. If noise is local, $H_{QCE}^{\rm ext}$ has the same locality as $H_{QC}$, i.e.,
\begin{equation}
H_{QCE}^{\rm ext} = \sum\limits_j {^j}\hspace{-.07cm}H_{QCE}^{\rm ext} \; ,
\end{equation}
\noindent where the support of ${^j}\hspace{-.07cm}H_{QCE}^{\rm ext}$ is contained in the union of the support of $H_{QC}^j$, the leakage space, and the environment. By repeating the same analysis as for local Hamiltonian noise, we find that local leakage noise satisfies eq.~(\ref{eq:local-noise-definition-pure}) with
\begin{equation}
\varepsilon = t_0 \cdot \max\limits_j ||{^j}\hspace{-.07cm}H_{QCE}^{\rm ext}||_{\infty} \; . 
\end{equation}
\vskip 0.0cm \HRule \vskip 0.3cm

\section{Encoded quantum computation} \vspace{.3cm}
\label{sec:encoded-quantum-computation}

When we desire to implement long computations---i.e., when the size $L$ of the quantum circuit is large---an accuracy that decreases linearly with $L$ as in eqs.~(\ref{eq:markovian-accuracy}) and (\ref{eq:non-markovian-accuracy}) is not satisfactory. Of course, we could achieve an accuracy independent of $L$ if $\varepsilon$ were a decreasing function of $L$ thus making $L\varepsilon$ a constant, but this is certainly not a physically reasonable assumption---we cannot hope that the hardware of the quantum computer will get less and less noisy the longer we keep quantum computing!  

In order to obtain the results of a quantum computation with constant accuracy, some method for detecting and correcting the errors that are introduced by the noisy hardware is necessary; we say that such a method of computation is {\em fault tolerant}. The basic idea of fault-tolerant computation, whether classical or quantum, is the use of {\em redundancy}: Every hardware operation in the circuit to be implemented is replaced by a collection of several hardware operations which are designed to be more robust to local noise than a single hardware operation alone. 

\begin{figure}[t]
\begin{tabular}{c}
\put(-4.2,0){\includegraphics[width=10cm,keepaspectratio,angle=-90]{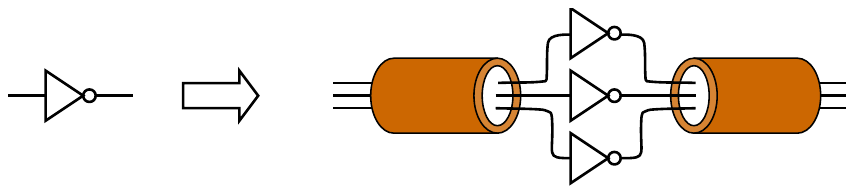}}
\vspace{-8.1cm}
\end{tabular}
\caption{\label{fig:classical-encoding} On the left, a {\sc not} gate is applied on the bit value carried in a single wire---e.g., if the {\sc not} operation is implemented by a {\sc cmos} inverter, the input wire controls the gate voltage and the output is taken as the drain voltage. The input and output wires on the left are replaced on the right by cables each comprising $n$ wires, and also the {\sc not} gate on the left is replaced on the right by an organ comprising $n$ parallel {\sc not} gates (here, $n=3$).}
\end{figure}

A formal method for introducing redundancy is via the use of error-correcting codes. 
For classical computation, the simplest example of a redundant encoding of information is based on the repetition code: ``To protect information traveling from gate to gate, we replace each wire of the noiseless circuit by a {\em cable} of $n$ wires (where $n$ is chosen appropriately); each wire within the cable is supposed to carry the same bit of information, and we hope that a majority will carry this bit even if some of the the wires fail.''\footnote[1]{Quote from Gacs's {\em Reliable computation}; see the references.} To protect information during the execution of each gate, we also replace each gate in the noiseless circuit by an {\em organ} comprising several gates. The organ operates on the information carried by the wires inside the cables in the same way that the initial unencoded gate operated in the information carried by single wires---e.g., a {\sc not} gate must be replaced by $n$ {\sc not} gates acting in parallel on every wire in a cable as in fig.~\ref{fig:classical-encoding}, and similarly a {\sc and} gate must be replaced by $n$ {\sc and} gates acting in parallel on corresponding pairs of wires in two cables. Organs also include a procedure for detecting and correcting faults in the noisy hardware; as discussed by von Neumann (see the references), this may be implemented for each wire by copying the value of every bit to a larger number of $k$ bits, randomly permuting all the resulting $n{\cdot}k$ bits, computing the majority function in parallel $n$ times on disjoint sets of $k$ of these permuted bits, and having the $n$ outputs form the output wire.

\begin{figure}[t]
\begin{tabular}{c}
\put(-7,0){\includegraphics[width=10cm,keepaspectratio,angle=-90]{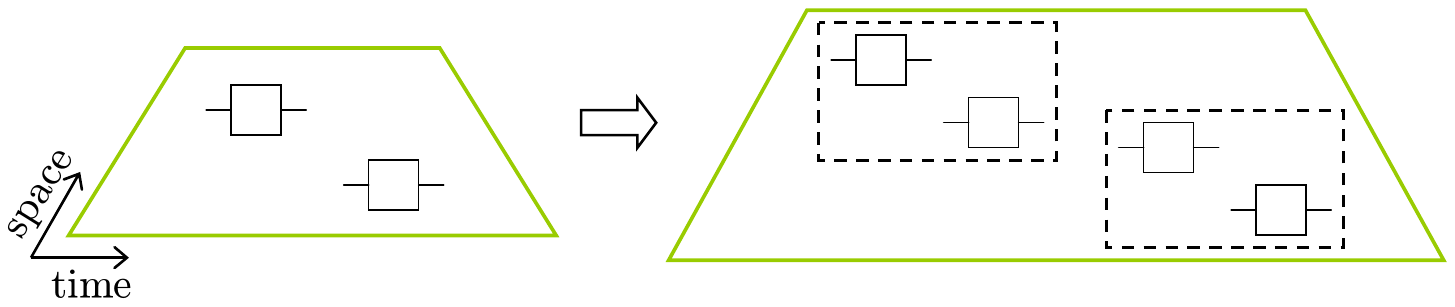}}
\vspace{-6.9cm}
\end{tabular}
\caption{\label{fig:quantum-encoding} Two elementary operations (single-qubit preparations, unitary gates, or measurements) in the noiseless quantum circuit on the left are replaced in the actual noisy quantum circuit on the right by two gadgets. For illustration, the elementary operations comprising each gadget are enclosed by dashed-line rectangles and only two elementary operations per gadget are shown.}
\end{figure}

Similar redundant encodings are also possible for quantum information where, for historical reasons, cables are now called {\em blocks} and organs are called {\em gadgets}\footnote[2]{People also use the term {\em rectangle} instead of {\em gadget} when they think of the latter's pictorial representation as in Fig.~\ref{fig:quantum-encoding}.}: Each qubit in the noiseless quantum circuit is replaced by a {\em block} of $n$ {\em encoded} qubits; the joint state of the encoded qubits is supposed to carry the same quantum information as the state of the initial qubit, and we hope that this quantum information can be recovered even if faults occur on some of the encoded qubits. In addition, each elementary operation in the noiseless quantum circuit is replaced by a {\em gadget} that comprises several elementary operations acting on the encoded qubits in a block or across multiple blocks. A gadget is designed to operate on the quantum information carried by the encoded qubits in the same way that the initial unencoded operation acted on the quantum information carried by single qubits, and it also includes a procedure for detecting and correcting faults in the noisy hardware. The encoding of quantum circuits is shown schematically in fig.~\ref{fig:quantum-encoding}.

But what do we mean when we say that the joint state of the encoded qubits carries the same quantum information as the state of the initial unencoded qubit? And how do we hope to recover the quantum information that is carried in a block if local noise acts on the encoded qubits? To answer the first question, consider a pure state $|\psi\rangle_{B_1R}$ supported on $\mathcal{H}_{B_1} \otimes \mathcal{H}_R$ where $\mathcal{H}_{B_1}$ is the Hilbert space of a qubit $B_1$ and $\mathcal{H}_R$ is the Hilbert space of a {\em reference} system $R$. The encoded version of $|\psi\rangle_{B_1R}$ is a pure state $|\psi\rangle_{BR}$ supported on $\mathcal{H}_n^0\otimes \mathcal{H}_R$, where $\mathcal{H}_n^0$ is the code space of a block $B$ of $n$ encoded qubits $B_1, \dots, B_n$. We say that the states $|\psi\rangle_{B_1R}$ and $|\psi\rangle_{BR}$ carry the same quantum information because $|\psi\rangle_{BR}$ can be obtained from $|\psi\rangle_{B_1R}$ by applying an isometry that maps $\mathcal{H}_{B_1}$ to $\mathcal{H}_n^0$; or in other words, there exists a unitary {\em decoding} unitary operator $U_{\rm dec}$ acting on the block such that
\begin{equation}
\label{eq:decoding-def}
{\rm Tr}_{(RB_1)^\perp} (U_{\rm dec} \otimes I_R) |\psi\rangle_{BR} = |\psi\rangle_{B_1R} \; ,
\end{equation} 
\noindent where $I_R$ is the identity operator on the reference system and ${\rm Tr}_{(RB_1)^\perp}$ denotes a trace over everything else except for the reference system and the qubit $B_1$.  

To answer the second question, we recall from Section \ref{sec:qc-discretization} that the basic idea of quantum error correction is error discretization. 
To monitor the effects of noise, we partition the entire Hilbert space of the encoded qubits into mutually orthogonal subspaces and, if there is no noise, we demand that the support of $|\psi\rangle_{BR}$ coincides at all times with the code space $\mathcal{H}_n^0$ (and the reference system). In the presence of noise, our strategy is to detect periodically  whether $|\psi\rangle_{BR}$ develops a non-zero overlap with any other subspace $\mathcal{H}_n^s$ labeled by a non-trivial syndrome $s$ (cf., eq.~(\ref{eq:discretize})), in which case we apply a recovery operation that returns the support of $|\psi\rangle_{BR}$ to $\mathcal{H}_n^0$. Physically, distinguishing on which subspace the state $|\psi\rangle_{BR}$ is supported can be implemented by performing a generalized measurement $\mathcal{M}_{\{\mu\}}$ jointly on the encoded qubits, processing the measurement outcome $\mu$ to determine the syndrome $s$ and hence the subspace $\mathcal{H}_n^s$ on which $|\psi\rangle_{BR}$ has been projected by the measurement\footnote[1]{This processing may be performed in a classical on-the-side computer.}, and applying an operation $\mathcal{R}_s$ on the encoded qubits conditioned on the value of $s$ that maps $\mathcal{H}_n^s$ to the code space $\mathcal{H}_n^0$.

We may revise eq.~(\ref{eq:decoding-def}) to include cases when the quantum information has been afflicted by noise: We now say that $|\psi\rangle_{B_1R}$ and the noisy $|\tilde\psi\rangle_{BR}$ carry the same quantum information if 
\begin{equation}
\label{eq:revised-definition-encoding}
{\rm Tr}_{(RB_1)^\perp} \left( \mathcal{D} (|\tilde\psi\rangle_{BR}) \right) = |\psi\rangle_{B_1R}  
\end{equation} 
\noindent with
\begin{equation}
\label{eq:decoder-definition}
\mathcal{D} = \mathcal{U}_{\rm dec} \circ \mathcal{R}_s \circ \mathcal{M}_{\{\mu\}} \otimes \mathcal{I}_R \; ,
\end{equation}
\noindent where $\mathcal{U}_{\rm dec}, \mathcal{I}_R$ are the physical operations corresponding to applying the unitaries $U_{\rm dec}, I_R$,  respectively, and we have suppressed the classical on-the-side computation which determines $s$ given $\mu$. It is worth of notice that the classical bits carrying the outcome $\mu$ of the syndrome measurement, which are traced over in eq.~(\ref{eq:revised-definition-encoding}), carry information about the subspace on which the quantum information was encoded prior to the decoding; for this reason, they are often referred to as {\em syndrome bits}.

The combined operation $\mathcal{D}$ is called a {\em decoder}. Because decoders output qubits which are unencoded and therefore unprotected from noise, we will {\em never} use a decoder in our actual noisy quantum circuits---encoded quantum information will never be decoded. A noiseless {\em ideal} decoder is however very useful as a tool for formalizing the requirement that a gadget operates on the encoded qubits in the same way that the operation that was replaced by the gadget acted on the initial unencoded qubits. 

\begin{figure}[t]
\begin{tabular}{c}
\put(-6.2,0){\includegraphics[width=12cm,keepaspectratio,angle=0]{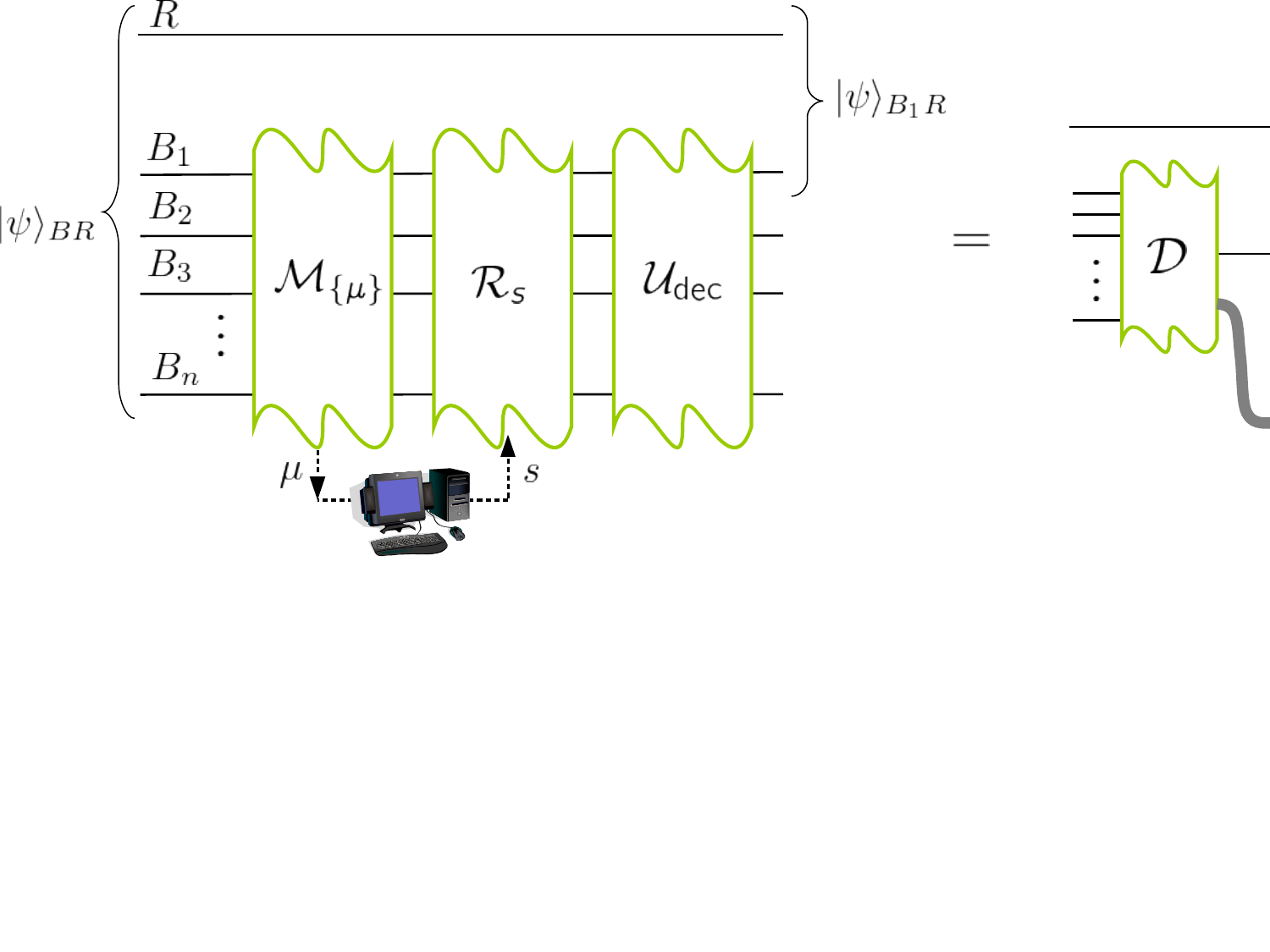}}
\vspace{-4cm}
\end{tabular}
\caption{\label{fig:ideal-decoder} A decoder $\mathcal{D}$ comprises a syndrome measurement $\mathcal{M}_{\{\mu\}}$, followed by a classical computation that outputs the syndrome $s$ with input the measurement outcome $\mu$, followed by a recovery operation $\mathcal{R}_s$ conditioned on the syndrome, followed by a decoding operation $\mathcal{U}_{\rm dec}$ mapping the code space of the block of encoded qubits $B_1, B_2, \dots, B_n$ to the Hilbert space of the qubit $B_1$. Our notation for $\mathcal{D}$ shows that the input consists of $n$ encoded qubits and the output of one decoded qubit, qubit $B_1$. As denoted by the bold gray line, the decoder output also includes the qubits $B_2, \dots, B_n$ and the {\em syndrome bits} that carry the outcome of the syndrome measurement; the state of these systems ends up being always a tensor product with the joint state of qubit $B_1$ and the reference $R$. When we speak of an {\em ideal} decoder, which is an imaginary noiseless operation, we draw $\mathcal{D}$ and the operations comprising it inside wavy boxes.}
\end{figure}

It is convenient to denote noiseless ideal operations using wavy boxes and noisy operations using square boxes; e.g., an {\em ideal} decoder is shown in fig.~\ref{fig:ideal-decoder}. With this notation, (a) an operation $\mathcal{P}_{|\psi\rangle}$ that prepares the single-qubit pure state $|\psi\rangle$ in the noiseless quantum circuit is replaced in the noisy quantum circuit by a gadget implementing the operation $\mathcal{P}_{|\psi\rangle}^L$ on the encoded qubits where \vspace{-.6cm}   
\begin{equation}
\label{eq:sim1}
\begin{tabular}{c}
\put(-3.55,0){\includegraphics[width=11cm,keepaspectratio,angle=0]{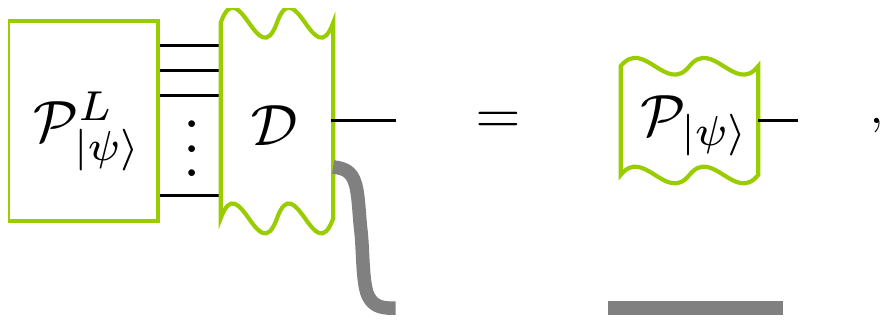}}
\vspace{-6.1cm}
\end{tabular}
\end{equation}
\noindent (b) an operation $\mathcal{U}$ that applies the single-qubit unitary $U$ is replaced by a gadget implementing the operation $\mathcal{U}_L$ on the encoded qubits where \vspace{-.6cm}
\begin{equation}
\label{eq:sim2}
\begin{tabular}{c}
\put(-3.65,0){\includegraphics[width=11cm,keepaspectratio,angle=0]{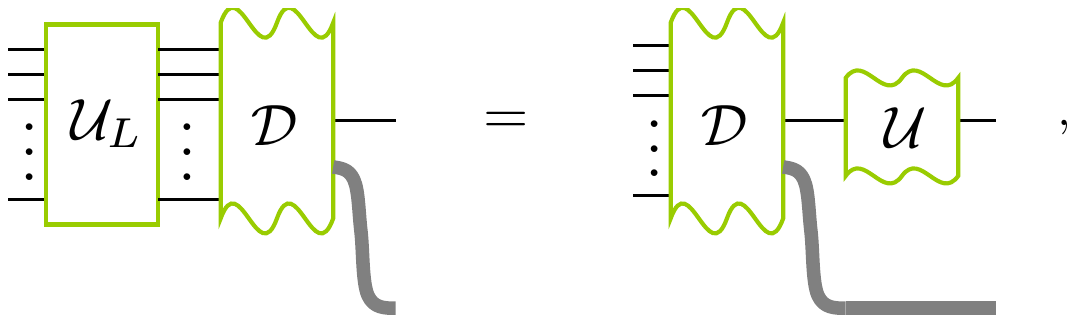}}
\vspace{-6.1cm}
\end{tabular}
\end{equation}

\noindent (and similarly for multi-qubit unitary operations), and (c) an operation $\mathcal{M}_{\{a\}}$ that measures a single-qubit observable $\hat a$ with eigenvalues $\{a\}$ is replaced by a gadget implementing the operation $\mathcal{M}_{\{a\}}^L$ on the encoded qubits where \vspace{-.6cm}
\begin{equation}
\label{eq:sim3}
\begin{tabular}{c}
\put(-2.6,0){\includegraphics[width=11cm,keepaspectratio,angle=0]{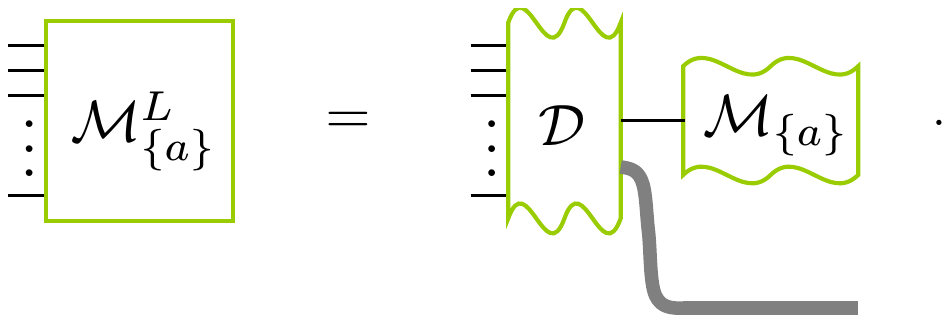}}
\vspace{-6.2cm}
\end{tabular}
\end{equation}

\noindent Eqs.~(\ref{eq:sim1}), (\ref{eq:sim2}), and (\ref{eq:sim3}) capture what we mean when we say that a gadget implementing the operation $\mathcal{O}_L$ {\em simulates} an unencoded operation $\mathcal{O}$. 

But apart from simulating the desired operation on the encoded qubits, gadgets need also contain a mechanism for detecting and correcting faults that may occur in the elementary operations comprising them. We can introduce such a mechanism by inserting an {\em error recovery} operation inside every gadget. Specifically, the gadget that implements $\mathcal{U}_L$ now becomes \vspace{-.6cm}   
\begin{equation}
\label{eq:sim4}
\begin{tabular}{c}
\put(-3.8,0){\includegraphics[width=11cm,keepaspectratio,angle=0]{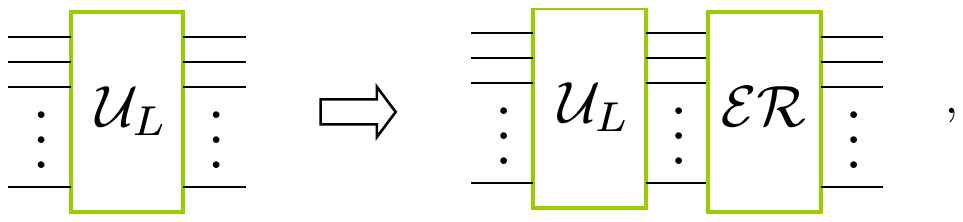}}
\vspace{-6.8cm}
\end{tabular}
\end{equation}
\noindent where $\mathcal{ER}$ comprises a syndrome measurement followed by a recovery operation, \vspace{-.6cm} 
\begin{equation}
\label{eq:sim5}
\begin{tabular}{c}
\put(-3.8,0){\includegraphics[width=11cm,keepaspectratio,angle=0]{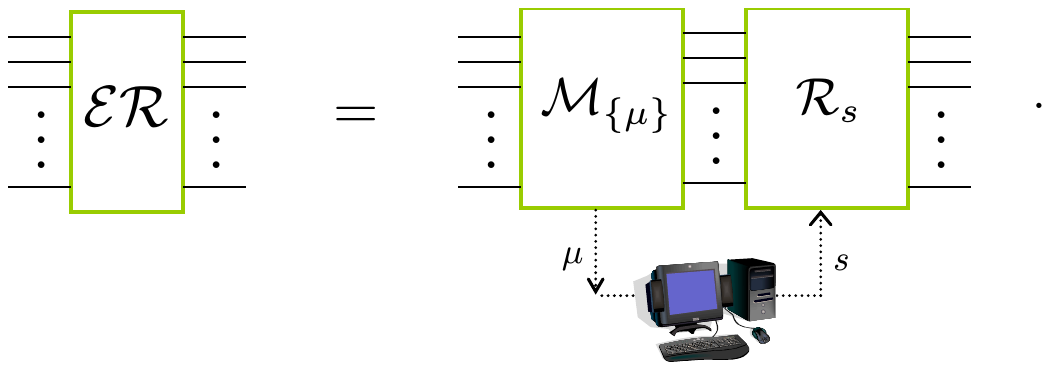}}
\vspace{-5.8cm}
\end{tabular}
\end{equation}

\noindent For the gadget that implements $\mathcal{P}_{|\psi\rangle}^L$, we insert similarly an error recovery operation following $\mathcal{P}_{|\psi\rangle}^L$, while the gadget that implements $\mathcal{M}_{\{a\}}^L$ is {\em not} modified since its output is a classical number, the measured eigenvalue $a$, which is protected from noise provided the classical computer that processes the measurement outcome is robust.

We do not have enough time and space to discuss explicit gadget constructions here. What is important at this abstract level is that each gadget simulates an operation in the noiseless quantum circuit, and that we hope that gadgets can be more robust to noise than unencoded operations because faults inside each gadget can be detected and corrected by the error recovery operation. 

\subsection{Properties of noisy gadgets} \vspace{.3cm}

To make progress, we need a method for formalizing the degree to which gadgets are protected from noise. We recall from Section \ref{sec:noise-models} that if the noisy quantum circuit is afflicted by local noise, we can expand the noisy evolution perturbatively as a sum of fault paths; each fault path identifies a specific subset of all the elementary operations where faults have been inserted, while there are no fault insertions in all remaining elementary operations. The idea of fault-tolerant constructions is to ensure that gadgets operate reliably for all the fault paths with no more than a certain number $t\,{>}\,1$ of insertions of faults inside them; the intuition is that each gadget will then be more robust to noise than any single elementary operation alone because the first contribution to a gadget's failure comes at order $t{+}1$ of our perturbative fault-path expansion.

\subsubsection{Good gadgets} \vspace{.3cm}
\label{sec:good-gadgets}

But what do we mean when we say that gadgets {\em operate reliably} for fault paths with at most $t$ faults inside them? To formalize this requirement we need to consider each gadget {\em together} with the error recovery operations of the immediately preceding gadgets; we will refer to a gadget together with its preceding error recovery operations as an {\em extended gadget}. The idea is to construct gadgets such that (a) for each fault path with at most $t$ insertions of faults inside a measurement extended gadget, the noisy gadget is equivalent to an ideal measurement of the gadget's ideally decoded input: \vspace{-.6cm}
\begin{equation}
\label{eq:correctness2}
\begin{tabular}{c}
\put(-3.8,0){\includegraphics[width=11cm,keepaspectratio,angle=0]{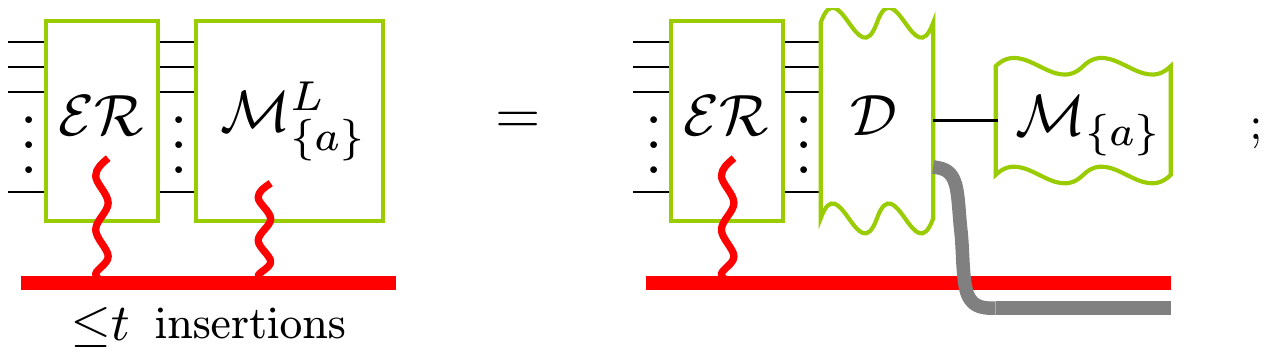}}
\vspace{-5.8cm}
\end{tabular}
\end{equation}
\noindent (b) for each fault path with at most $t$ insertions of faults inside an extended gadget simulating a single-qubit unitary gate, the gadget is equivalent to applying the ideal unitary gate to the gadget's ideally decoded input\footnote[1]{And similarly for multi-qubit unitary gates where we include all preceding error recovery operations.}: \vspace{-.4cm}
\begin{equation}
\label{eq:correctness}
\begin{tabular}{c}
\put(-5.7,0){\includegraphics[width=11cm,keepaspectratio,angle=0]{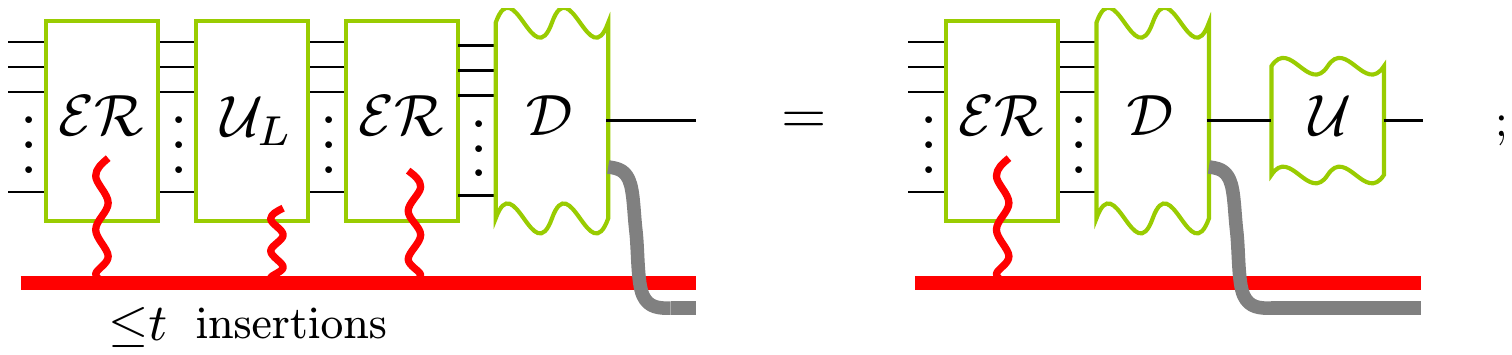}}
\vspace{-5.9cm}
\end{tabular}
\end{equation}
\noindent and (c) for each fault path with at most $t$ insertions of faults inside a preparation gadget\footnote[2]{The extended preparation gadget coincides with the preparation gadget since preparation gadgets have no input and, therefore, they are not preceded by any error recovery operations.}, the gadget is equivalent to the ideal preparation: \vspace{-.5cm}
\begin{equation}
\label{eq:correctness3}
\begin{tabular}{c}
\put(-4.8,0){\includegraphics[width=11cm,keepaspectratio,angle=0]{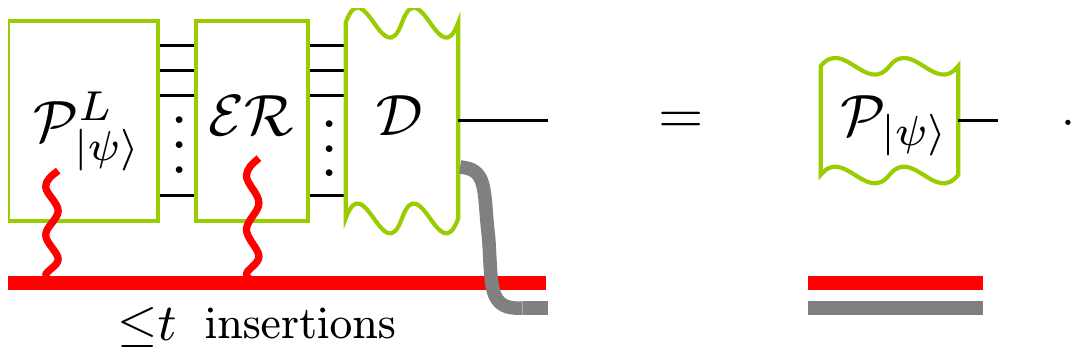}}
\vspace{-5.8cm}
\end{tabular}
\end{equation}
\noindent Here, we have illustrated insertions of faults as couplings between the noisy elementary operations inside the gadgets and the environment denoted by a think red line. 

Figuratively, eq.~(\ref{eq:correctness2}) allows us to {\em create} ideal decoders out of measurement extended gadgets which contain at most $t$ faults, eq.~(\ref{eq:correctness}) allows us to {\em propagate} ideal decoders to the left through unitary-gate extended gadgets which contain at most $t$ faults, and eq.~(\ref{eq:correctness3}) allows us to {\em annihilate} ideal decoders in preparation gadgets which contain at most $t$ faults. As ideal decoders are created in measurement gadgets, propagated through unitary-gate gadgets, and annihilated in preparation gadgets, the noisy encoded quantum circuit is transformed to a noiseless {\em unencoded} quantum circuit. 

We conclude that for all the fault paths with at most $t$ faults in each and every gadget, a noisy quantum computer is equivalent to a noiseless ideal quantum computer, in the sense that they both produce the same probability distribution for the final measurements which determine the computation result. However, for fault paths for which there are more than $t$ faults inside one of the extended gadgets, the output probability distributions from the noisy and the noiseless quantum computer are not guaranteed to agree; thus it appears that the accuracy $1\,{-}\,\delta$ of a noisy {\em encoded} quantum circuit composed of fault-tolerant gadgets scales with
\begin{equation}
\label{eq:accuracy-claim}
\delta \sim \varepsilon^{t+1} \; ,
\end{equation}
\noindent which should be compared with the scaling $\delta\,{\sim}\,\varepsilon$ for a noisy {\em unencoded} quantum circuit---cf., eq.~(\ref{eq:markovian-accuracy}) for local Markovian noise and eq.~(\ref{eq:non-markovian-accuracy}) for local non-Markovian noise. To prove that this scaling indeed holds and to determine the proportionality constant in eq.~(\ref{eq:accuracy-claim}) requires that we analyze what happens when a gadget is afflicted by more than $t$ faults.

\subsubsection{Bad gadgets} \vspace{.3cm}
\label{sec:bad-gadgets}

Eqs.~(\ref{eq:correctness2}), (\ref{eq:correctness}), and (\ref{eq:correctness3}) show that extended gadgets that contain at most $t$ faults can be viewed as implementing ideal operations acting on unencoded qubits. But if more than $t$ fault occur inside a gadget, there are no guarantees what may happen. Although we cannot in general say much about what {\em actually} happens unless we know  details about the noise and how gadgets are constructed, we will be satisfied if we can show that extended gadgets which contain more than $t$ faults can be viewed as implementing some {\em noisy} operations acting on unencoded qubits.

\begin{figure}[t]
\begin{tabular}{c}
\put(-6.2,0){\includegraphics[width=12cm,keepaspectratio,angle=0]{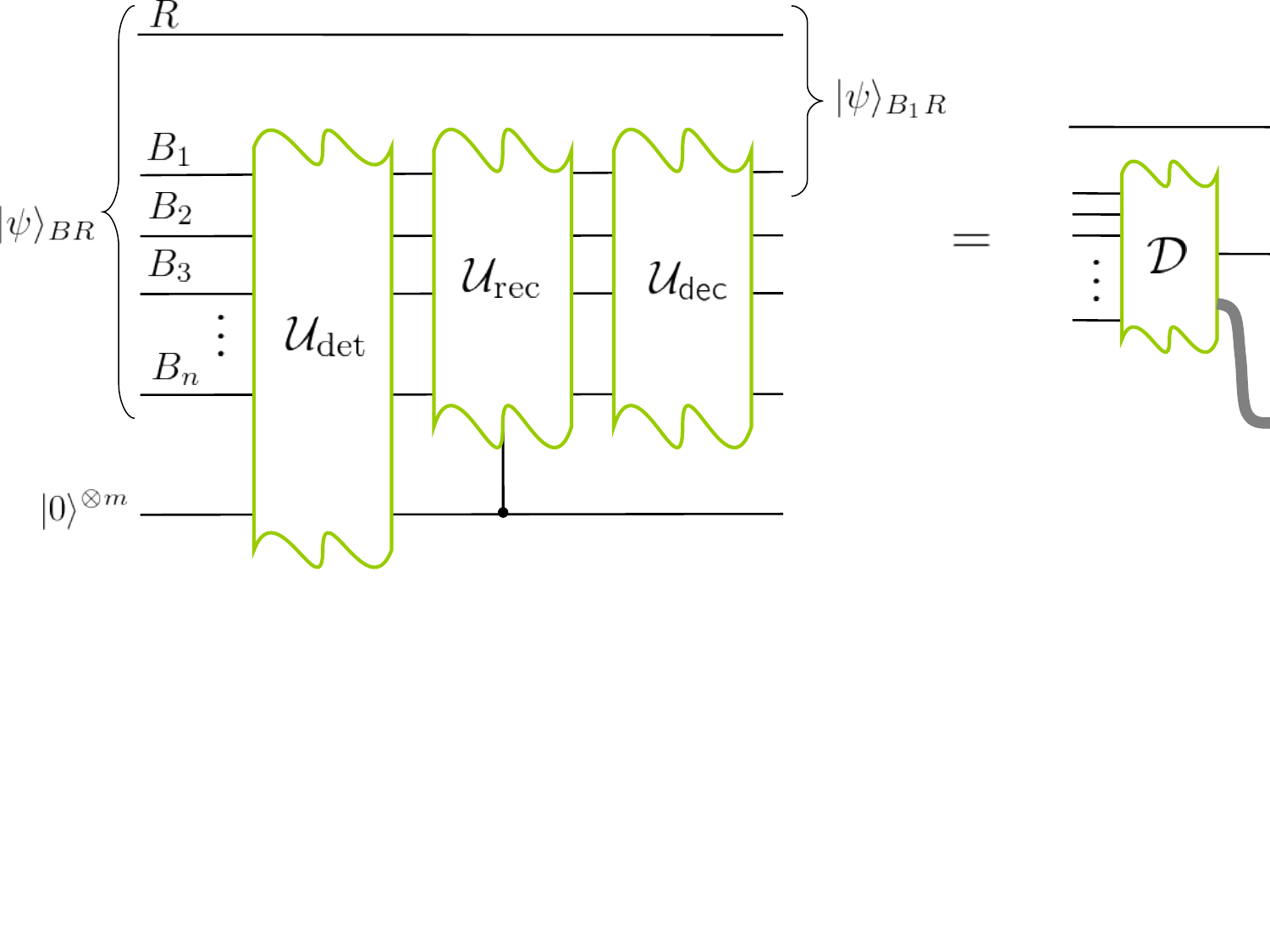}}
\vspace{-3.7cm}
\end{tabular}
\caption{\label{fig:reversible-ideal-decoder} A reversible ideal decoder, to be contrasted with the irreversible ideal decoder in fig.~\ref{fig:ideal-decoder}. First, the syndrome measurement $\mathcal{M}_{\{\mu\}}$ is replaced by a unitary operation $\mathcal{U}_{\rm det}$ that acts on a larger Hilbert space which includes ancillary {\em syndrome qubits}; this unitary maps coherently the state of the syndrome qubits to a state corresponding to the subspace $s$ on which the input block is supported. Secondly, the recovery operation $\mathcal{R}_s$ is replaced by a controlled unitary operation $\mathcal{U}_{\rm rec}$ that applies the appropriate recovery unitary on the block depending on the syndrome $s$ carried by the syndrome qubits. In the figure, we assume there are $m$ syndrome qubits and that each one is prepared in the state $|0\rangle$.} 
\end{figure}

The basic tool we will need is a decomposition of the identity, i.e. do nothing, operation in terms of an ideal {\em decoder-encoder pair}$\,$: \vspace{-.6cm} 
\begin{equation}
\label{eq:encoder-decoder-pair}
\begin{tabular}{c}
\put(-3.2,0){\includegraphics[width=11cm,keepaspectratio,angle=0]{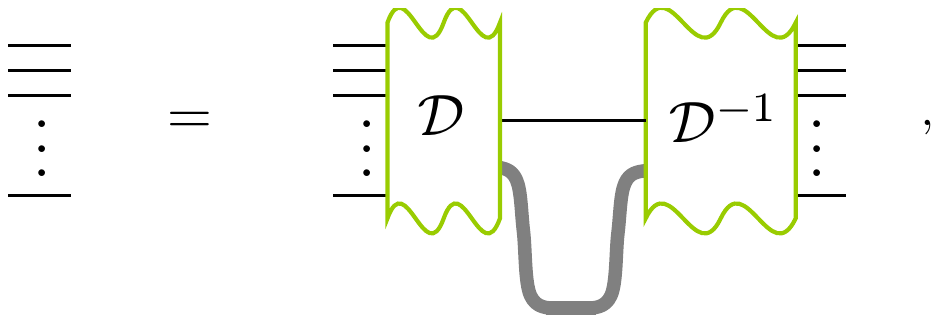}}
\vspace{-6.1cm}
\end{tabular}
\end{equation}
\noindent where the operation $\mathcal{D}^{-1}$ is an {\em ideal encoder}. Of course, for ideal decoders to have an inverse, they need to be implemented by a reversible circuit. In fact, we can easily modify our definition of ideal decoders in Section \ref{sec:encoded-quantum-computation} to achieve reversibility. Fig.~\ref{fig:reversible-ideal-decoder} shows such a reversible ideal decoder; an ideal encoder can then be implemented by simply executing the circuit in this figure backward in time.

Decoder-encoder pairs are trivial (they multiply to the identity operation) but they can be useful for understanding the properties of our noisy circuits if we insert them strategically at appropriate places. As a start, consider how to modify eq.~(\ref{eq:correctness}) in the case there are more than $t$ faults. The following property now holds: For each fault path with more than $t$ insertions of faults inside an extended gadget simulating a single-qubit unitary gate, the gadget is equivalent to applying some noisy gate to the extended gadget's ideally decoded input: \vspace{-.5cm}
\begin{equation}
\label{eq:badness}
\begin{tabular}{c}
\put(-5.5,0){\includegraphics[width=11cm,keepaspectratio,angle=0]{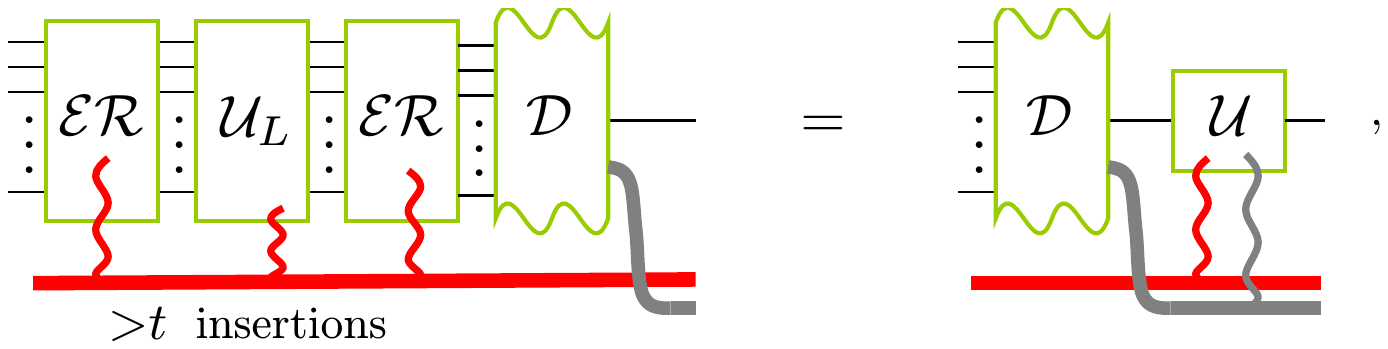}}
\vspace{-5.8cm}
\end{tabular}
\end{equation}
\noindent where the noisy gate is \vspace{-.6cm}
\begin{equation}
\label{eq:encoder-decoder}
\begin{tabular}{c}
\put(-4.2,0){\includegraphics[width=11cm,keepaspectratio,angle=0]{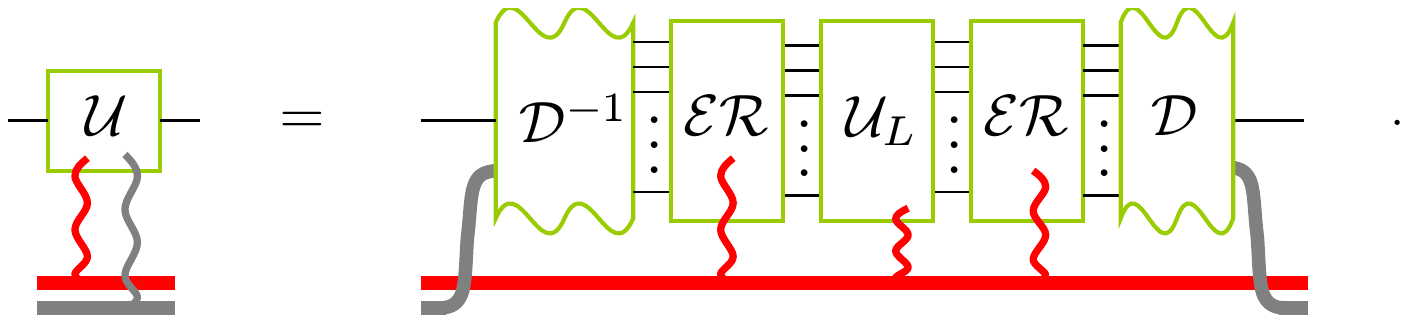}}
\vspace{-5.8cm}
\end{tabular}
\end{equation}

There are two points about eq.~(\ref{eq:badness}) deserving emphasis. First, this property replaces the entire extended gadget by a (noisy) gate rather than replacing the gadget alone; cf. eq.~(\ref{eq:correctness}) and see Section \ref{sec:truncation} below. Secondly, the noisy gate that replaces the extended gadget depends on both the faults inside the extended gadget and also on the syndrome bits (found inside the bold gray lines) which are input to the ideal encoder. As our notation in eq.~(\ref{eq:encoder-decoder}) is intended to illustrate, the syndrome bits can be viewed as a fictitious environment which operates together with the actual environment associated with the noise. 

Similar properties as eq.~(\ref{eq:badness}) can be derived for measurement extended gadgets and preparation gadgets; schematically, \vspace{-.5cm}
\begin{equation}
\label{eq:badness3}
\begin{tabular}{c}
\put(-3.7,0){\includegraphics[width=11cm,keepaspectratio,angle=0]{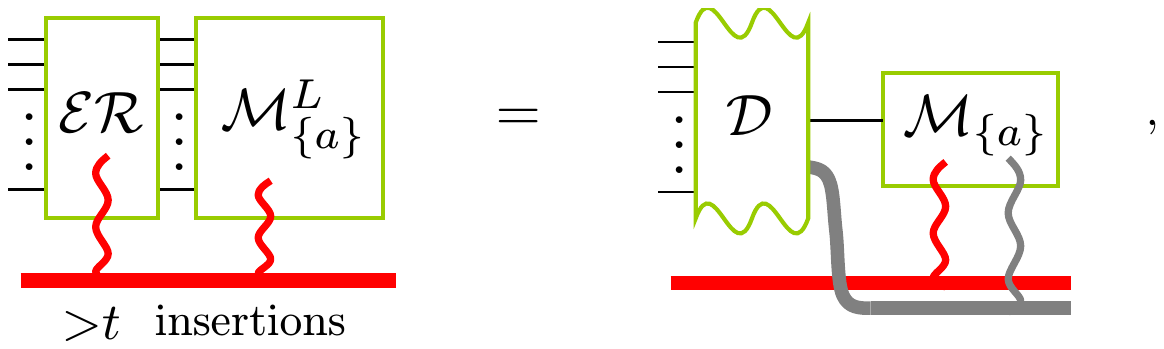}}
\vspace{-5.8cm}
\end{tabular}
\end{equation}
\noindent where the noisy measurement on the right-hand side can be obtained by inserting a decoder-encoder pair on the left-hand side, and \vspace{-.6cm}
\begin{equation}
\label{eq:badness2}
\begin{tabular}{c}
\put(-4.9,0){\includegraphics[width=11cm,keepaspectratio,angle=0]{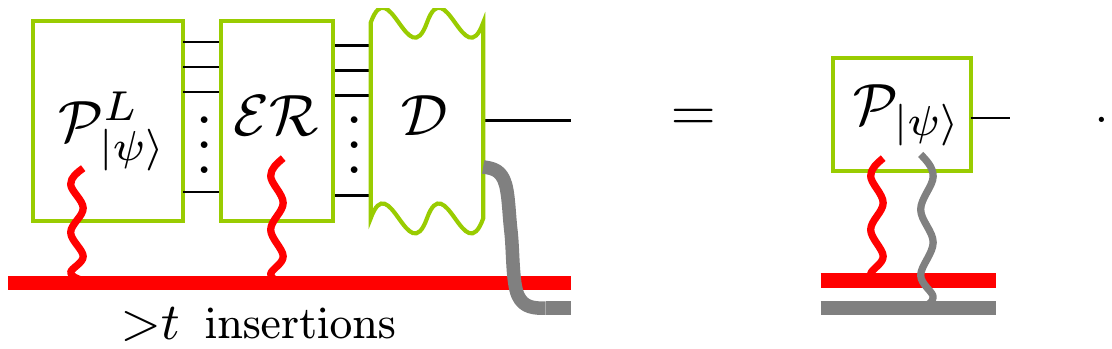}}
\vspace{-5.8cm}
\end{tabular}
\end{equation}

\subsubsection{Truncation} \vspace{.3cm}
\label{sec:truncation}

Figuratively, eqs.~(\ref{eq:badness}), (\ref{eq:badness3}), and (\ref{eq:badness2}) allow us to create, to propagate to earlier times, and to annihilate ideal decoders in the case when extended gadgets are {\em bad} containing more than $t$ faults. These properties are therefore complementary to eqs.~(\ref{eq:correctness2}), (\ref{eq:correctness}), and (\ref{eq:correctness3}) that apply to {\em good} extended gadgets containing at most $t$ faults. 

Of course, one difference between the two cases is that in one case noisy unencoded operations appear on the right-hand side, while in the other case the unencoded operations are noiseless and ideal. But here we would like to discuss a second difference that was mentioned already in the previous section; namely, while eqs.~(\ref{eq:badness}), (\ref{eq:badness3}), and (\ref{eq:badness2}) replace the {\em entire} bad  extended gadgets by some (noisy) unencoded operations, eqs.~(\ref{eq:correctness2}), (\ref{eq:correctness}), and (\ref{eq:correctness3}) only replace the {\em gadgets} contained inside the good extended gadgets by the ideal unencoded operations. This modification is important in order to prevent overcounting faults in successive, and therefore overlapping, bad extended gadgets. 

To be concrete, imagine that we are to encode a quantum circuit comprising just a single-qubit preparation $\mathcal{P}_{|\psi\rangle}$ followed by a single-qubit measurement $\mathcal{M}_{\{a\}}$. The encoded quantum circuit comprises two overlapping extended gadgets: \vspace{-.5cm}
\begin{equation}
\label{eq:truncation}
\begin{tabular}{c}
\put(-4.3,0){\includegraphics[width=11.2cm,keepaspectratio,angle=0]{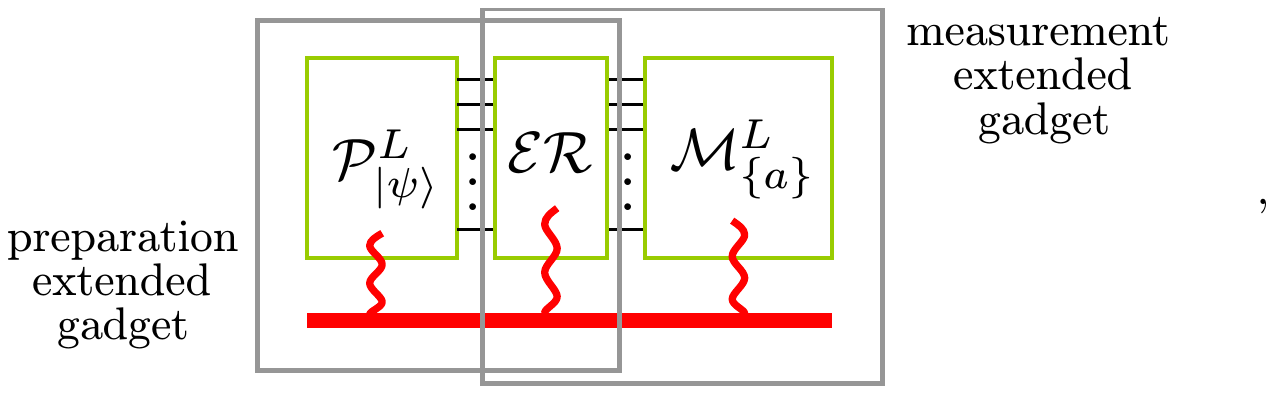}}
\vspace{-5.7cm}
\end{tabular}
\end{equation}
\noindent where the error recovery step is contained in both extended gadgets enclosed in the square gray boxes. Now, consider a fault path with more than $t$ insertions of faults in each extended gadget. Considering the two extended gadgets in isolation, we may think that the noisy encoded circuit equals some noisy preparation followed by some noisy measurement: \vspace{-.5cm}
\begin{equation}
\label{eq:truncation2}
\begin{tabular}{c}
\put(-4.8,0){\includegraphics[width=11.2cm,keepaspectratio,angle=0]{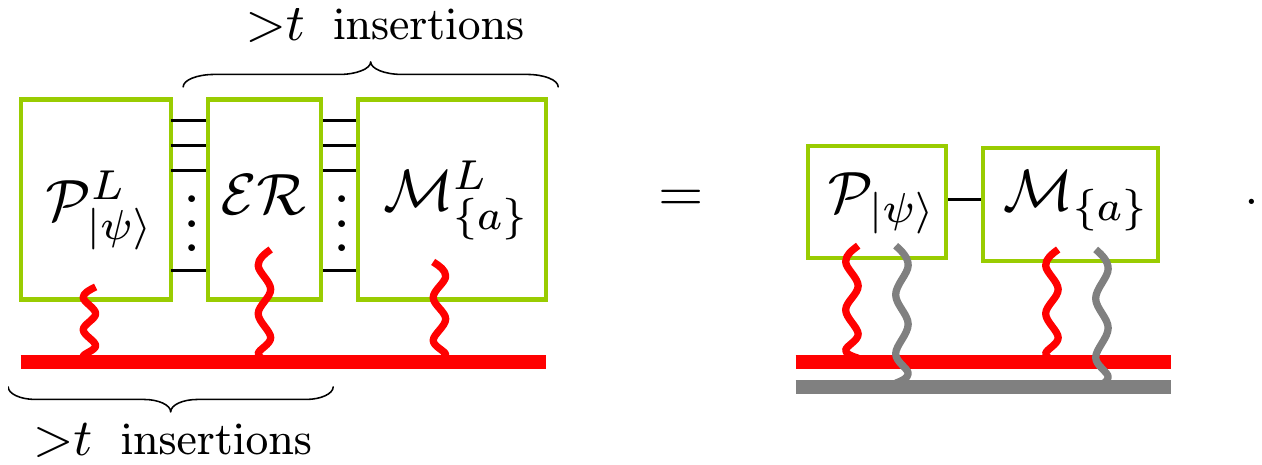}}
\vspace{-5.2cm}
\end{tabular}
\end{equation}
\noindent However, we soon realize that this may not always be a satisfying answer. Because the two extended gadgets overlap, it is possible that the {\em total} number $N_f$ of faults in the given fault path is less than $2(t{+}1)$; thus the two noisy unencoded operations on the right-hand side appear at order $\varepsilon^{N_f}$ in our perturbative fault path expansion, which is less than the order $\left( \varepsilon^{t+1} \right)^2$ we would expect based on the fact that each extended gadget by itself fails at order $\varepsilon^{t+1}$. Clearly, the problem is that our naive estimate $\left( \varepsilon^{t+1} \right)^2$ double counts each fault inside the error recovery step shared by the two overlapping extended gadgets. 

Nevertheless, this complication is a red herring, and our naive estimate can actually be justified: To formally obtain eq.~(\ref{eq:truncation2}), we first need to use eq.~(\ref{eq:badness3}) thereby replacing the entire measurement extended rectangle by a noisy unencoded measurement: \vspace{-.6cm}
\begin{equation}
\label{eq:truncation4}
\begin{tabular}{c}
\put(-4.8,0){\includegraphics[width=11.2cm,keepaspectratio,angle=0]{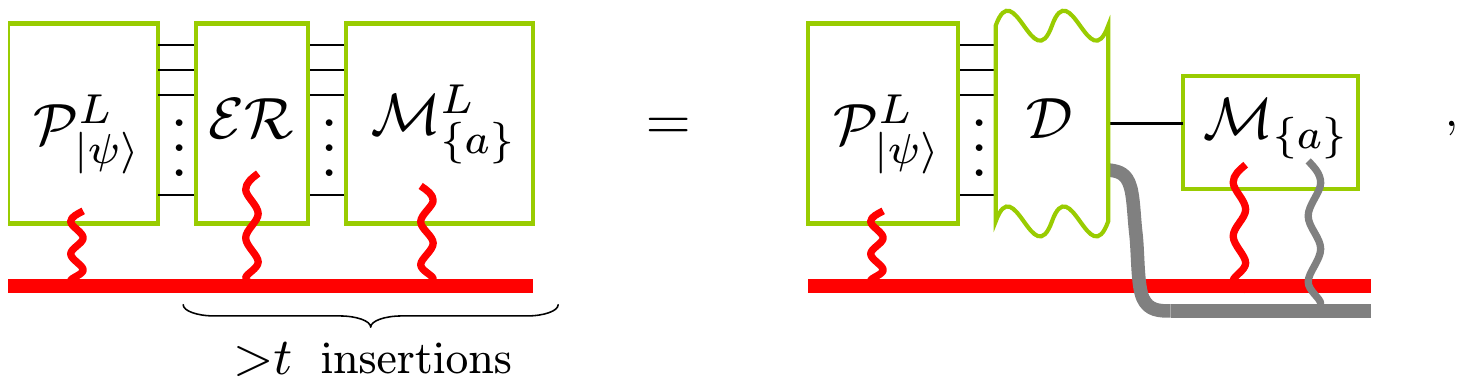}}
\vspace{-5.7cm}
\end{tabular}
\end{equation}

\noindent where we observe that the error recovery step has been removed from the preparation extended gadget---we say that the gadget has been {\em truncated}. To annihilate the ideal decoder at the next step, we need to consider how many faults are contained inside the {\em truncated} preparation gadget; if there are at most $t$ faults then we can use eq.~(\ref{eq:correctness3}) to obtain \vspace{-.6cm}
\begin{equation}
\label{eq:truncation3}
\begin{tabular}{c}
\put(-4.8,0){\includegraphics[width=11.2cm,keepaspectratio,angle=0]{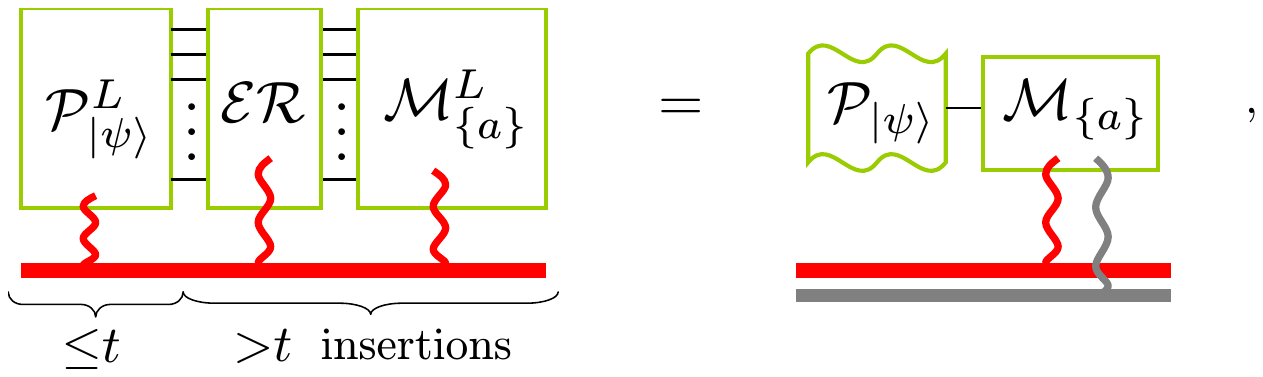}}
\vspace{-6cm}
\end{tabular}
\end{equation}
\noindent whereas, if there are more than $t$ faults, we can use eq.~(\ref{eq:badness2}) to obtain eq.~(\ref{eq:truncation2}); see, however, the next paragraph which explains why eq.~(\ref{eq:badness2}) applies to truncated extended gadgets. Since in this second step what matters is the number of faults in the truncated preparation gadget, there is no double counting of faults; thus the two noisy unencoded operations on the right-hand side of eq.~(\ref{eq:truncation2}) do appear at order $\left( \varepsilon^{t+1} \right)^2$ in our perturbative fault path expansion, as desired. 

We need to note that eqs.~(\ref{eq:correctness2}), (\ref{eq:correctness}), and (\ref{eq:correctness3}) for good extended gadgets and eqs.~(\ref{eq:badness}), (\ref{eq:badness3}), and (\ref{eq:badness2}) for bad extended gadgets, which have been formulated for the full extended gadgets, apply in the same way to truncated extended gadgets. Indeed, the ideal decoders contain a noiseless ideal error recovery step---cf. fig.~(\ref{fig:ideal-decoder})---which can be used to replace the truncated noisy error recovery steps, thereby reassembling the full gadgets for which the properties apply.

Although we have illustrated the concept of truncation with a simple example, a similar truncation procedure can be used for any fault path which leads to more than two successive bad extended gadgets, including gadgets that simulate unencoded operations on more than one qubit: Starting from the latest bad extended gadgets which are not succeeded by any other bad extended gadgets, we progressively move to all earlier bad extended gadgets, one gadget at a time. At each step, we truncate the bad extended gadget under consideration from the error recovery steps it shares with all its succeeding bad extended gadgets (which may themselves be truncated or not), and we label the truncated gadget as good or bad depending on the number of faults it contains after the truncation (if it contains at most $t$ faults, it is declared good, otherwise it is declared bad). Eventually we reach the earliest bad extended gadgets, and we truncate the good extended gadgets that preceded them\footnote[1]{These extended gadgets remain good after the truncation as they were already good prior to it.}. In the end, the successive bad extended gadgets are divided into {\em non-overlapping} truncated extended gadgets which have been declared good or bad depending on the number of faults they contain and which can be replaced by either noisy or ideal unencoded operations respectively by using the corresponding gadget properties.  

\section{Coarse-grained noise and level reduction} \vspace{.3cm}
\label{sec:level-reduction}

We have now assembled all the properties we need to characterize noisy gadgets. If we combine all the pieces together, we can arrive at a helpful description of the noise acting on the encoded quantum computer. 

We recall from Section \ref{sec:noise-models} that, without encoding, the only fault path leading to the noiseless ideal evolution is the trivial fault path which contains absolutely no faults; we are then forced to decompose the noisy evolution into an ideal and a faulty part as in eq.~(\ref{eq:fault-path-expansion}) for local Markovian noise and eq.~(\ref{eq:fault-path-expansion-pure}) for local non-Markovian noise. In contrast, in an encoded quantum computation, many more fault paths lead to the noiseless ideal evolution; now, for local Markovian noise, we can decompose the noisy evolution as  
\begin{equation}
\label{eq:good-bad}
\rho^{\rm noisy} = \rho^{\rm good} + \zeta^{\rm bad} \; ,
\end{equation}
\noindent and similarly for local non-Markovian noise,
\begin{equation}
\label{eq:good-bad-pure}
|\psi\rangle^{\rm noisy} = |\psi\rangle^{\rm good} + |\vartheta\rangle^{\rm bad} \; ,
\end{equation}
\noindent where the unnormalized density matrix $\rho^{\rm good}$ and the unnormalized pure state $|\psi\rangle^{\rm good}$ are sums of all the fault paths with at most $t$ faults in each and every extended gadget, while $\zeta^{\rm bad}$ and $|\vartheta\rangle^{\rm bad}$ sum of all the remaining fault paths.

In what sense are the fault paths included in $\zeta^{\rm good}$ and $|\vartheta\rangle^{\rm good}$ good? For each ({\em good}) fault path with at most $t$ faults in each and every extended gadget, eqs.~(\ref{eq:correctness2}), (\ref{eq:correctness}), and (\ref{eq:correctness3}) apply. We now consider the entire encoded quantum circuit and, by using these properties, we first create ideal decoders in the measurement gadgets. Then, we propagate the ideal decoders to earlier times through unitary-gate gadgets. Finally, we annihilate the ideal decoders in preparation gadgets. As the ideal decoders appear, move to earlier times, and finally disappear, the entire encoded quantum circuit afflicted by the given good fault path is shown to be formally equal to the ideal quantum circuit that the gadgets simulate. Since this is true for every good fault path separately, by linearity it is also true for the sum of all of them; thus $\zeta^{\rm good}$ and $|\vartheta\rangle^{\rm good}$ lead (after normalization) to the same probability distribution for the final computation result as a noiseless ideal quantum computer would.

We can now estimate the accuracy $1\,{-}\,\delta$ of the encoded quantum computation. For local Markovian noise, $\delta$ can be bounded as in eq.~(\ref{eq:base-accuracy}) by the norm of the difference of the final noisy superoperator minus its (normalized) good part:
\begin{equation}
\label{eq:encoded-accuracy-markovian}
\delta \leq ||\rho^{\rm noisy} - {\rho^{\rm good} \over {\rm Tr}\left(\rho^{\rm good}\right) } ||_1 \leq \left( 1 + {1\over 1 - || \zeta^{\rm bad} ||_1} \right) || \zeta^{\rm bad} ||_1 \; ,
\end{equation}
\noindent and we have used the triangle inequality multiple times\footnote[2]{We have $\delta \leq || \alpha \rho^{\rm good} + \zeta^{\rm bad} ||_1 \leq |\alpha| + || \zeta^{\rm bad} ||_1 $, where $|\alpha| = {\rm Tr}^{-1}{\left(\rho^{\rm good}\right)} - 1 \leq || \zeta^{\rm bad} ||_1 / (1 - || \zeta^{\rm bad} ||_1)$.}. Similarly, for local non-Markovian noise, $\delta$ can be bounded as in eq.~(\ref{eq:base-accuracy-non-Markovian}) by the norm of the difference of the final noisy pure quantum state minus its (normalized) good part:
\begin{equation}
\label{eq:encoded-accuracy-non-markovian}
\delta \leq 2 || |\psi\rangle^{\rm noisy} - {|\psi\rangle^{\rm good} \over || |\psi\rangle^{\rm good} ||} || \leq 2 \left( 1 + {1\over 1 - || |\vartheta\rangle^{\rm bad} || } \right) || |\vartheta\rangle^{\rm bad} || \; .
\end{equation}

It remains to obtain upper bounds on $\zeta^{\rm bad}$ and $|\vartheta\rangle^{\rm bad}$ which are sums of all ({\em bad}) fault paths with more than $t$ faults in at least one gadget. For each bad fault path, the noisy encoded quantum circuit can be analyzed by using the gadget properties: We first consider whether each extended gadget contains at most $t$ faults or more than $t$ faults, declaring the former gadgets good and the latter bad. If multiple successive extended gadgets are declared bad, we use the truncation procedure described in Section~(\ref{sec:truncation}) to divide them into non-overlapping truncated extended gadgets which are good or bad depending on the number of faults they contain. Eventually, we use eqs.~(\ref{eq:badness}), (\ref{eq:badness3}), and (\ref{eq:badness2}) for the good extended gadgets (truncated or not) and eqs.~(\ref{eq:correctness2}), (\ref{eq:correctness}), and (\ref{eq:correctness3}) for the bad extended gadgets (also, truncated or not). Every good extended gadget is thereby replaced by the noiseless ideal operation the gadget simulates, every bad extended gadget is replaced by some noisy operation, and thus the noisy encoded quantum circuit as a whole is replaced by a noisy unencoded quantum circuit.

If we now let $C^{(1)}$ denote the set of all $L$ extended gadgets in the encoded quantum circuit, then by analogy to eq.~(\ref{eq:fault-paths}) we may write
\begin{equation}
\label{eq:inclusion-exclusion-markovian}
\zeta^{\rm bad} = \sum\limits_{r=1}^L (-1)^{r-1} \sum\limits_{C^{(1)}_r\subseteq C^{(1)}} \zeta( C^{(1)}_r ) \; ,
\end{equation} 
\noindent where the second sum is over all subsets $C^{(1)}_r$ of $C^{(1)}$ of cardinality $r$, and $\zeta( C^{(1)}_r )$ denotes a sum of all the fault paths for which all the extended gadgets in $C^{(1)}_r$ are declared bad\footnote[1]{Whether each extended gadget in $C^{(1)}_r$ is truncated depends on the fault path; however, for each specific fault path, we can first use the truncation procedure to decide which extended gadgets need to be truncated, and then we can unambiguously declare every extended gadget (truncated or not) as being either good or bad.}. 

To gain intuition about how to proceed, consider the simplest case $r\,{=}\,1$ when $\zeta( C^{(1)}_1 )$ is a sum of all the fault paths for which the single extended gadget in a {\em specific} set $C^{(1)}_1$ is bad. Since for an extended gadget (truncated or not) to be bad it needs to contain more than $t$ faults, we can generalize eq.~(\ref{eq:fault-paths}) to obtain
\begin{equation}
\label{eq:inclusion-exclusion-example}
\zeta( C^{(1)}_1 ) = \sum\limits_{s=t+1}^{L_0} (-1)^{s-t-1} {s{-}1 \choose t} \sum\limits_{C_s} \zeta( C_s ) \; ,
\end{equation} 
\noindent where $L_0$ is the number of elementary operations in the extended gadget in $C^{(1)}_1$, the second sum is over all subsets $C_s$ of $s$ of these operations, and $\zeta( C_s )$ is a sum of all the fault paths with faults applied on all operations in $C_s$ \footnote[2]{We first sum all $\zeta( C_{t{+}1} )$ accounting correctly for all the fault paths with exactly $t{+}1$ faults; however, all the fault paths with exactly $t{+}2$ faults are overcounted ${t{+}2 \choose t{+}1}{-}1 = {t{+}1\choose t}$ times. So next, we subtract the sum of all $\zeta( C_{t{+}2} )$ multiplied by ${t{+}1\choose t}$, but in doing so we undercount all the fault paths with exactly $t{+}3$ faults ${t{+}3\choose t{+}1}{-}{t{+}1\choose t} {t{+}3\choose t{+}2}{-}1 = -{t{+}3{-}1\choose t}$ times. And so on.}. 

For the general case of $r$ bad extended gadgets, it suffices to perform a similar inclusion-exclusion analysis independently in each gadget:
\begin{equation}
\label{eq:inclusion-exclusion-general}
\zeta( C^{(1)}_r ) = \prod\limits_{j=1}^{r} \left( \sum\limits_{s_j=t+1}^{L_0} (-1)^{s_j-t-1} {s_j{-}1 \choose t} \sum\limits_{C_{s_j} \subseteq C^{(1)}_r(j) } \zeta( C_{s_j} ) \right) \; ,
\end{equation} 
\noindent where $L_0$ now denotes the number of elementary operations in the {\em largest} extended gadget\footnote[3]{The number of elementary operations may vary among gadgets. In addition, some gadgets may be truncated depending on the fault path. By taking $L_0$ to correspond to the largest extended gadget, we thus unavoidably include in the sum some extra fault paths which should not be counted. However, since eventually we will take the norm of both sides and use the triangle inequality, including these additional fault paths merely weakens our bounds.}, $C^{(1)}_r(j)$ denotes the set of elementary operations in the $j$-th bad extended gadget in $C^{(1)}_r$, and $C_{s_j}$ denotes a subset of $s_j$ of the elementary operations in $C^{(1)}_r(j)$. 

But $||\zeta( C_{s_j} ) ||_1 \leq \varepsilon^{s_j}$ by the definition of local Markovian noise, and we find
\begin{equation}
\label{eq:level-reduction-markovian} 
||\zeta( C^{(1)}_r )||_1 \leq  \prod\limits_{j=1}^{r} \sum\limits_{s_j=t+1}^{L_0} {s_j{-}1 \choose t} {L_0 \choose s_j} \varepsilon^{s_j} \leq \left( {L_0 \choose t{+}1} \varepsilon^{t+1} \sum\limits_{\omega=0}^{\infty} {(L_0{-}t{-}1)^\omega \varepsilon^\omega \over \omega!} \right)^r \leq \left( \varepsilon^{(1)} \right)^r \;  ,
\end{equation} 
\noindent with
\begin{equation}
\label{eq:renormalized-strength-markovian}
\varepsilon^{(1)} = \xi {L_0 \choose t{+}1} \varepsilon^{t+1} \; 
\end{equation} 
\noindent for some constant $\xi\geq e^{(L_0{-}t{-}1) \varepsilon}$ (typically we are interested in small values $\varepsilon \leq 1/(L_0{-}t{-}1)$; then we may take $\xi$ to be $e$). By replacing $\zeta$ with $|\vartheta\rangle$, we may repeat a similar calculation for local non-Markovian noise to obtain 
\begin{equation}
\label{eq:level-reduction-non-markovian} 
|||\vartheta( C^{(1)}_r )\rangle|| \leq \left( \varepsilon^{(1)} \right)^r \; , 
\end{equation} 
\noindent with $\varepsilon^{(1)}$ again as in eq.~(\ref{eq:renormalized-strength-markovian}).

\begin{figure}[t]
\begin{tabular}{c}
\put(-6.3,0){\includegraphics[width=13cm,keepaspectratio]{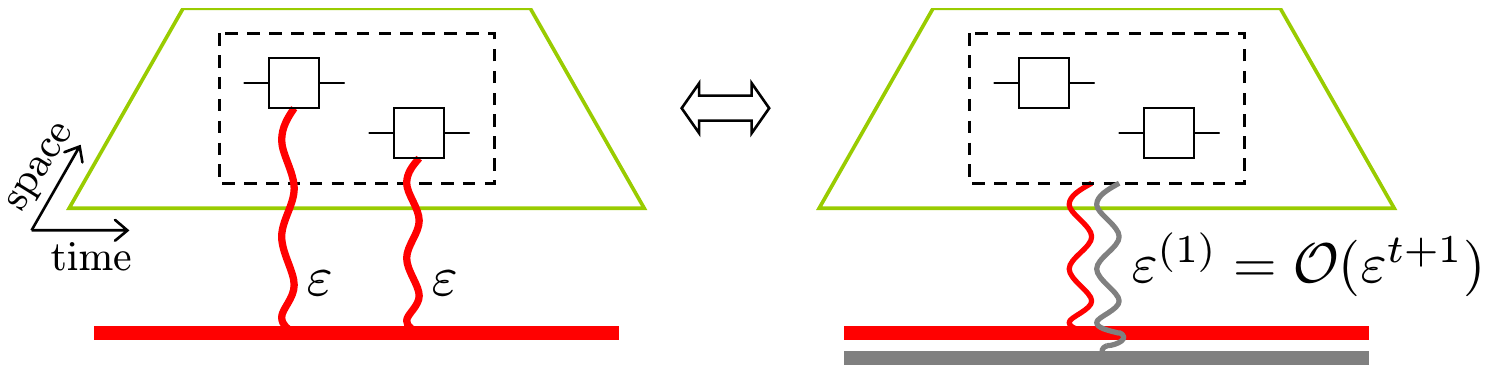}}
\vspace{-6.5cm}
\end{tabular}
\caption{\label{fig:coarsegraining} The encoded operations in an encoded quantum circuit are executed by using gadgets (here, one gadget is shown, along with two of the elementary operations it contains); the elementary operations inside each gadget are afflicted by local noise (either Markovian or non-Markovian) of strength $\varepsilon$. The physical noise can be coarse-grained into a local effective noise with renormalized strength $\varepsilon^{(1)}$ of order $\varepsilon^{t+1}$ acting on the gadgets themselves. The noise coarse-graining allows us to concentrate on the gadgets alone and forget about the elementary operations inside them---we say that the encoded quantum circuit is {\em level reduced} to an equivalent unencoded quantum circuit, where the effect of coding is to map the physical noise strength $\varepsilon$ to the effective noise strength $\varepsilon^{(1)}$.}
\end{figure}

Eqs.~(\ref{eq:level-reduction-markovian}) and (\ref{eq:level-reduction-non-markovian}) tell us that if we choose any $r$ extended gadgets, then the sum of all the fault paths for which all of them are bad has norm which is exponentially suppressed with $r$. This is exactly the condition we imposed for noise to be local, except this in this case we think of noise as afflicting the gadgets themselves instead of the elementary operations. While the strength of the noise acting on the elementary operations is $\varepsilon$, the strength of the coarse-grained noise acting on the gadgets is $\varepsilon^{(1)}$, which scales as $\varepsilon^{t+1}$ because each gadget can tolerate up to $t$ faults. We often refer to the encoded computation executed by the gadgets as a {\em level-1} simulation of an unencoded {\em level-0} quantum circuit; in this language, what we have shown is that a noisy level-1 simulation afflicted by local noise with strength $\varepsilon$ can be viewed as a {\em level reduced} noisy level-0 simulation afflicted by a coarse-grained local noise with renormalized strength $\varepsilon^{(1)}$ (and the level reduction works the same for both Markovian and non-Markovian noise). Fig.~\ref{fig:coarsegraining} illustrates this coarse-graining level reduction procedure.

The accuracy of an encoded computation afflicted by local Markovian noise can now be determined from eq.~(\ref{eq:encoded-accuracy-markovian}) by combining eqs.~(\ref{eq:inclusion-exclusion-markovian}) and (\ref{eq:level-reduction-markovian}); we find  
\begin{equation}
\label{eq:delta-level-1-circuit}
\delta \approx ||\zeta^{\rm bad}||_1 \leq \sum\limits_{r=1}^L {L\choose r} \left( \varepsilon^{(1)} \right)^r \leq (e{-}1)L\varepsilon^{(1)} \; ,
\end{equation} 
\noindent where, in the first step, we have kept only the leading order contribution. By comparing with eqs.~(\ref{eq:markovian-accuracy}) which corresponds to the case no encoding is used, we conclude that if $\varepsilon^{(1)} < \varepsilon$ then the encoding is in fact a good idea since it improves the accuracy $1\,{-}\,\delta$ of the final computation result. The same conclusion also holds for local non-Markovian noise.

\section{The quantum accuracy threshold} \vspace{.3cm}
\label{sec:threshold-theorem}

We are one breath away from the central result in the theory of quantum fault tolerance. You must have guessed the next step... If an encoded quantum circuit is more accurate than a quantum circuit which is not encoded, then why not apply the encoding to the encoded circuit itself, taking every elementary operation inside it and replacing it by a gadget; this doubly encoded quantum circuit should be even more accurate. In fact, why stop here? If we continue recursively re-encoding our encoded circuits, we expect their accuracy to steadily increase reaching any limit we please. 

\begin{figure}[t]
\begin{tabular}{c}
\put(-4.7,0){\includegraphics[width=13cm,keepaspectratio]{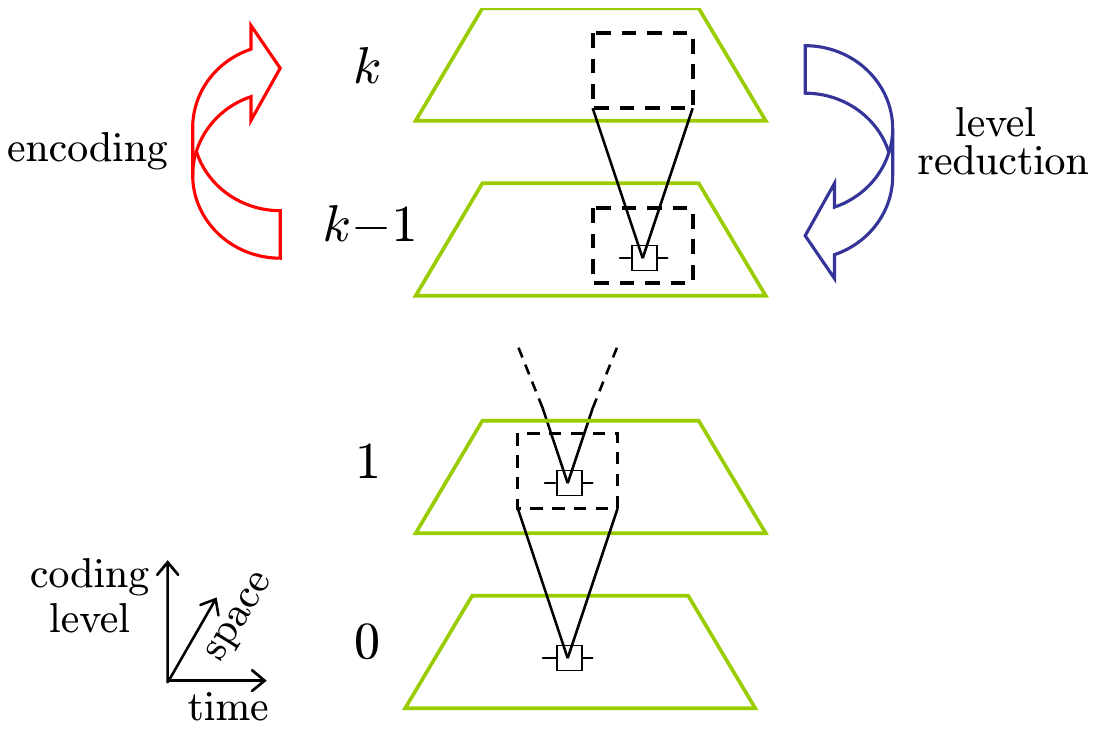}}
\vspace{-3.7cm}
\end{tabular}
\caption{\label{fig:concatenation} A recursive construction of a multiply encoded quantum circuit. At the base, the unencoded level-0 circuit corresponds to our quantum algorithm. One level higher, the encoded level-1 circuit is obtained by simulating every elementary operation in the level-0 circuit by using a gadget. By repeating this replacement rule, we eventually obtain an encoded level-$k$ circuit which is the circuit we physically implement. To understand the effect of local noise on the level-$k$ circuit, we can level reduce it to a level-$(k{-}1)$ circuit which is acted by noise of renormalized strength. If level reduction is repeated $k$ times, the level-$k$ circuit can be level reduced to an unencoded level-0 circuit; we can view the strength of noise acting on this level-0 circuit as the {\em effective} noise strength acting on the multiply encoded operations in the level-$k$ circuit.}
\end{figure}

To formalize this idea, let us consider the recursive construction of these multiply encoded quantum circuits we imagined above. At the base of our construction is the unencoded quantum circuit corresponding to our quantum algorithm; we say that this is our level-0 circuit in the sense that it does not use any coding. The next step is to replace every elementary operation in the level-0 circuit by the corresponding gadget; we say that this is our level-1 circuit, performing a level-1 simulation of the level-0 circuit. Instead of physically implementing the level-1 circuit as is, we may next replace every elementary operation in the level-1 circuit by the corresponding gadget to obtain our level-2 circuit, and so on. Fig.~\ref{fig:concatenation} illustrates this replacement procedure repeated $k$ times; the final encoded quantum circuit, which is the one we {\em do} physically implement, performs a level-$k$ simulation of the level-0 circuit. 

The question is what is the accuracy of the level-$k$ circuit as a function of $k$. Estimating this accuracy is actually especially easy if we use the noise coarse-graining concept from Section \ref{sec:level-reduction}: The noise afflicting the elementary operations in the level-$k$ circuit can be coarse-grained to give an effective noise that acts on the gadgets; if the physical noise is local and has strength $\varepsilon$, the coarse-grained noise is also local and has renormalized strength $\varepsilon^{(1)}$ as in eq.~(\ref{eq:renormalized-strength-markovian}). The noise coarse-graining level reduces the level-$k$ circuit to an equivalent level-$(k{-}1)$ circuit; this level-$(k{-}1)$ circuit produces the same probability distribution for the computation outcome as the initial level-$k$ circuit, but it is afflicted by an effective local noise of strength $\varepsilon^{(1)}$. Thus we have reduced the problem of estimating the accuracy of the level-$k$ circuit (the circuit we actually physically implement) to estimating the accuracy of the level-reduced level-$(k{-}1)$ circuit (which is, of course, imaginary as it represents the result of our noise coarse-graining procedure).

We can next coarse grain the noise in the level-$(k{-}1)$ circuit, thereby level reducing the initial level-$k$ circuit to a level-$(k{-}2)$ circuit afflicted by local noise of strength $\varepsilon^{(2)}$; because of the self-similarity of our recursive circuit construction, the map from $\varepsilon^{(1)}$ to $\varepsilon^{(2)}$ is the same as from $\varepsilon$ to $\varepsilon^{(1)}$. The noise in the level-$(k{-}2)$ circuit can in turn be coarse grained, thereby level reducing the initial level-$k$ circuit to a level-$(k{-}3)$ circuit. And so on, where at the $l$-th coarse-graining step the input noise strength $\varepsilon^{(l-1)}$ is renormalized to an output strength 
\begin{equation}
\label{eq:l-th-level-reduction}
\varepsilon^{(l)} \leq \xi {L_0 \choose t{+}1} \left( \varepsilon^{(l-1)}\right)^{t+1} \; .
\end{equation} 

After $k$ coarse-graining steps, the initial level-$k$ circuit is eventually level reduced to an {\em unencoded} level-0 circuit  afflicted by local noise of strength $\varepsilon^{(k)}$---this level-0 circuit corresponds to the quantum algorithm which is simulated by the initial encoded level-$k$ circuit since every level reduction takes us down one level in the ladder in fig.~(\ref{fig:concatenation}). If we use the recursion eqs.~(\ref{eq:l-th-level-reduction}) where $\varepsilon^{(0)}=\varepsilon$ is the physical noise strength, we find 
\begin{equation}
\label{eq:level-k-noise-strength}
\varepsilon^{(k)} \leq \varepsilon_0 \left( {\varepsilon \over \varepsilon_0} \right)^{(t+1)^k} \; , \; {\rm for \,\, a \,\, constant} \;\; \varepsilon_0 = \left( \xi {L_0 \choose t{+}1} \right)^{-1/t} \; .
\end{equation} 
\noindent The constant $\varepsilon_0$ is the critical noise strength below which the recursive encoding scheme we have described is effective; if $\varepsilon\,{<}\,\varepsilon_0$ then $\varepsilon^{(k)}$ decreases double exponentially with the coding level $k$. The critical noise strength $\varepsilon_0$ is often referred to as the {\em threshold} for fault-tolerant quantum computation.

The $k$ successive level reduction steps tell us that we can view the encoded operations in the level-$k$ circuit as being afflicted by an effective local noise of strength $\varepsilon^{(k)}$. Thus, the accuracy $1\,{-}\,\delta$ of the level-$k$ circuit can be estimated as in Section \ref{sec:level-reduction} where we estimated the accuracy of a level-1 circuit afflicted by local noise of strength $\varepsilon^{(1)}$. For local Markovian noise, we now have
\begin{equation}
\delta \lesssim (e{-}1)L\varepsilon^{(k)} \; ,
\end{equation}
\noindent where we used eq.~(\ref{eq:delta-level-1-circuit}) with $\varepsilon^{(1)}$ replaced by $\varepsilon^{(k)}$. If the physical noise has a strength below the threshold, $\varepsilon\,{<}\,\varepsilon_0$, then $\delta$ can become as small as desired by recursively re-encoding the level-0 circuit of our quantum algorithm sufficiently many times $k$. The same conclusion also holds for local non-Markovian noise.

In particular, imagine that we desire to obtain the computation output with an accuracy $1\,{-}\,\delta \geq 1\,{-}\,\delta_0$ for some constant (error) $\delta_0$, independent of the size $L$ of the quantum algorithm. We can arrange to have $\delta \leq \delta_0$ by choosing $k$ so that
\begin{equation}
\label{eq:overhead-1}
(t{+}1)^k \leq {\log\left( (e{-}1)L\varepsilon_0 / \delta_0 \right) \over \log\left( \varepsilon_0 / \varepsilon \right)} \; .
\end{equation}
\noindent Because of the recursiveness of our encoding construction, each of the $L$ encoded operations in the level-$k$ circuit can be implemented by using at most $(L_0)^k$ elementary operations, where $L_0$ is the number of elementary operations in the largest gadget. The ratio then of the number $L^*$ of elementary operations in the entire level-$k$ circuit over the number $L$ of elementary operations in the quantum algorithm scales as
\begin{equation}
\label{eq:overhead-2}
{L^*\over L} \leq \left(L_0\right)^k = {\it O} \left(\log L\right)^a \; , \; {\rm with} \; \; a = {\log L_0 \over \log(t{+}1)} \; .
\end{equation}
\noindent Thus, not only does the level-$k$ circuit achieve the desired accuracy $1\,{-}\,\delta_0$, but it does so very efficiently; the level-$k$ circuit is only larger than the unencoded level-0 circuit by a polynomial in the {\em logarithm} of the size $L$ of the quantum algorithm. 

\section{Assessment} \vspace{.3cm}
\label{sec:review}

The idea in the previous section was to establish the existence of a critical noise strength by considering a specific fault tolerance scheme; in particular, we chose to study the recursive scheme illustrated in fig.~\ref{fig:concatenation} because its self-similar nature greatly simplified our analysis. Although recursive schemes are easier to analyze, it is clearly possible that they are not optimal from a practical point of view, and other more complex schemes may have higher thresholds and/or more favorable overhead costs. Proposing and analyzing improved schemes for fault-tolerant quantum computation is a major focus of current research.

The existence of a critical noise strength is significant because it implies that the quest to build a reliable quantum computer is not a mere fantasy, but it is based on firm foundations: We have learned that if we find a physical setting allowing us to experimentally implement quantum circuits with local noise of strength $\varepsilon_0$ or less, then the noisy operations can be assembled efficiently to perform an encoded quantum computation and obtain the computation result to as high an accuracy as desired. 

Although this knowledge gives us confidence and encouragement to research further how quantum computers can be constructed, it is possible that the outcome of this endeavor may ultimately be failure. We can contemplate several possibilities under which such a failure might occur: First, it is possible that the entire concept of what it means to quantum compute, a concept which is based on the laws of conventional quantum mechanics, is flawed when applied to quantum computers with either a very large number of qubits or very long running times---clearly, in all our considerations we have assumed (as the majority of physicists currently believe) that the framework of quantum mechanics can be extrapolated without change to the (long) time and (large) length scales relevant for quantum computers implementing useful computations. Perhaps there are fundamental, as yet unknown, principles that prevent the realization of the highly entangled multi-particle quantum states required to implement useful quantum algorithms. In this sense, the project of quantum computing can be seen under a different light, the light of testing quantum mechanics in new regions of the parameter space; even if nothing useful as regards computation comes out, we may uncover puzzles forcing us to revise our approach to quantum mechanics and physics in general.

A second possibility for failure relates to the conditions we imposed on the noise as we formulated our theoretical analysis. It is possible that, as we design and test quantum computing devices of increasing complexity, we will eventually find that the physical noise is not captured by our local noise models or that, even if noise is local, its strength cannot be upper bounded by a small constant number. 
Ultimately, the question of whether methods of quantum fault tolerance can in practice be as effective as our theoretical analysis indicates will be decided by the progress of the future experiments. In the mean time, theoretical research has still ample room for further progress: Fruitful new research can attempt to relax the requirements under which reliable quantum computation can be provably shown to be possible; e.g., one may consider more precise models for the noise during qubit preparation and measurement, one may specialize to noise models that more closely describe the particular characteristics of observed decoherence in modern prototype experimental devices, etc. 

At present, we have no evidence neither that quantum mechanics is violated at the length and time scales relevant for long useful quantum computations, nor that the physical noise in prospective implementations of quantum computation has features that prevent quantum fault tolerance from working. Certainly, there are formidable technical difficulties for building a large-scale quantum computer with present technology, and it is possible that the engineering requirements may prove too challenging to overcome for a long time in the future. Nevertheless, experimental efforts during the last decade have shown great progress, and there is a great sense of optimism among experimentalists that this progress will continue even more rapidly as they gain more insight and intuition about their systems.  

\section{History and further reading} \vspace{.3cm}
\label{sec:history}

\vspace{.2cm}
\hspace{3cm}\parbox{12cm}{ \footnotesize
{\it Computer engineering is the art and science of translating user requirements we do not fully understand; into hardware and software we cannot precisely analyze; to operate in environments we cannot accurately predict; all in such a way that the society at large is given no reason to suspect the extent of our ignorance. }

--- adapted from Kaplan's {\em By Design: Why There Are No Locks on the Bathroom Doors in the Hotel Louis XIV and Other Object Lessons}, Fairchild Books (2004). 
} \vspace{.8cm}

In the hope of making the flow of thought in this chapter as smooth as possible, we have avoided interruptions to discuss the history of the subject of quantum fault tolerance and we have also omitted discussing a number of technical but important details. This final section provides some of this historical context and references to published works---most of which are available freely on the {\em arXiv.org} servers---where further information can be obtained. Of course, knowing that our historical account and our list of references cannot be perfectly complete, our aspiration is not to provide an exhaustive list of all relevant publications but rather to guide the interested reader in his/her first steps in the large bibliography. 

The question whether logical operations can be implemented fault tolerantly despite noise was central from the early days of the development of classical computing. Shannon's master's thesis \cite{Shannon40} laid the foundations of digital circuit design, and von Neumann's analysis of noisy cellular automata \cite{Neumann55,Neumann66} showed how unreliable components can be assembled to implement reliable computations. A more recent exposition of methods for reliable classical computation can be found in Gacs' work \cite{Gacs05}; see also Gacs' works \cite{Gacs83,Gacs01} on noisy cellular automata and Gray's guide \cite{Gray01} on \cite{Gacs83}. 

The corresponding study for quantum computing was pioneered by Shor \cite{Shor96} who described the first gadget constructions for universal quantum computation. Soon after, the existence of a critical noise strength based on a recursive scheme as in fig.~\ref{fig:concatenation} was discussed by Aharonov and Ben-Or \cite{Aharonov96,Aharonov99}, by Kitaev \cite{Kitaev97,Kitaev97b}, and by Knill, Laflamme, and Zurek \cite{Knill96b}. All these works considered local Markovian noise and made a series of additional assumptions about the experimental quantum computing devices. Most notably, one assumes that one can supply fresh ancillary qubits or refresh existing qubits at any point in time during the noisy quantum computation (this is a necessary assumption; if it is dropped, there is no critical noise strength \cite{Aharonov96b}). In addition, one assumes that there is maximum parallelism, i.e., it is possible to apply gates in parallel on disjoint sets of qubits (also a necessary assumption \cite{Preskill97}). Finally, one assumes that multi-qubit gates can be applied between any set of qubits irrespective of their geometric distance (this is not a necessary assumption; a critical noise strength exists even when geometric constraints are taken into account \cite{Gottesman00}). 

More recently, a new proof for the existence of a critical noise strength was described by Aliferis, Gottesman, and Preskill \cite{Aliferis05b}; see also Aliferis' doctoral thesis \cite{Aliferis07a} and Gottesman's review \cite{Gottesman09}. This proof is significantly simpler than earlier proofs, and it applies to both Markovian and non-Markovian local noise (the analysis for non-Markovian local noise extends prior results by Terhal and Burkard \cite{Terhal04}). Building on this new proof, Aharonov, Kitaev, and Preskill \cite{Aharonov05} later analyzed long-range static noise, Aliferis and Terhal \cite{Aliferis05c} analyzed leakage noise, Aliferis and Preskill \cite{Aliferis07d} analyzed biased noise with dephasing being much more dominant than relaxation or leakage, and Ng and Preskill \cite{Preskill08} analyzed Gaussian noise.

The value of the critical noise strength has been estimated both analytically by means of combinatorial analyses and also, for simple probabilistic local noise models, by performing numerical simulations in a classical computer. The highest numerical estimates to date (of order $1.0\times 10^{-2}$) have been obtained for Knill's postselection and Fibonacci schemes \cite{Knill05} and for Raussendorf, Harrington, and Goyal's scheme based on surface codes \cite{Raussen07b}. The highest analytical estimates to date (of order $1.0\times 10^{-3}$) have been obtained by Reichardt \cite{Reichardt06b}, by Aliferis, Gottesman, and Preskill \cite{Aliferis07b}, and by Aliferis and Preskill \cite{Aliferis08}, all by analyzing Knill's schemes and modifications of them. It is interesting to note that the schemes with the highest known critical noise strengths share two features: First, they make use of quantum teleportation \cite{Bennett93} for implementing quantum error correction \cite{Knill05} and for simulating certain gates \cite{Gottesman99}. Secondly, they use a method by Bravyi and Kitaev \cite{Bravyi04} for {\em distilling} high accuracy copies of certain ancillary quantum states out of noisier copies of the same states.   

\vskip 0.0cm \HRule \vskip 0.1cm
\hskip 11cm -- {\em Panos Aliferis} (2009)

\hskip 10.4cm {\em panos@alumni.caltech.edu}


\bibliographystyle{alpha}

\tableofcontents

\end{document}